\documentclass[cameraready]{usetex-v1}

\usepackage{graphicx}
\usepackage{times}
\usepackage{cite}	
\usepackage{url}
\usepackage{xspace}
\usepackage{textcomp}

\long\def\com#1{}

\long\def\xxx#1{}

\long\def\abbr#1#2{#1}			

\newcommand{\abcite}[2]{\abbr{\cite{#1}}{\cite{#1,#2}}}

\newcommand{\blind}[1]{}		

\renewenvironment{itemize}{
   \begin{list}{\labelitemi}{
     \setlength{\topsep}{0.5ex}
     \setlength{\itemsep}{-0pt}
     \setlength{\itemindent}{0pt}
     \setlength{\leftmargin}{\labelwidth}
     \addtolength{\leftmargin}{-8pt}}
}{\end{list}}

\renewenvironment{enumerate}{
   \begin{list}{\labelenumi}{
     \usecounter{enumi}
     \setlength{\topsep}{0.5ex}
     \setlength{\itemsep}{-0pt}
     \setlength{\itemindent}{0pt}
     \setlength{\leftmargin}{\labelwidth}
     \addtolength{\leftmargin}{-2pt}}
}{\end{list}}

\let\origdescription\description
\renewenvironment{description}{
  \setlength{\topsep}{0.5ex}
  \setlength{\itemsep}{-0pt}
  \setlength{\leftmargini}{0pt}
  \origdescription
  \setlength{\itemindent}{0pt}
}
{\endlist}

\def\tcpoo{$u$TCP\xspace}
\def\utcp{$u$TCP\xspace}
\def\mstcp{$ms$TCP\xspace}

\def\ucobs{$u$COBS\xspace}

\def\utls{$u$TLS\xspace}

\begin{document}
\title{\vspace{-1em}
  {\LARGE Fitting Square Pegs Through Round Pipes} \\
	\vspace{.2em}
	{\it Unordered Delivery
	Wire-Compatible with TCP and TLS}
}



\author{
        Michael F. Nowlan$^\dag$ ~~ Nabin Tiwari$^*$ ~~
        Janardhan Iyengar\thanks{Franklin and Marshall College\abbr{}{, Email: \{jiyengar, nabin.tiwari\}@fandm.edu}} ~~
        Syed Obaid Amin$^\dag$ ~~
        Bryan Ford\thanks{Yale University\abbr{}{, Email: \{bryan.ford, michael.nowlan\}@yale.edu}}
}
\maketitle

\begin{abstract}

Internet applications increasingly employ TCP
not as a {\em stream abstraction},
but as a {\em substrate} for application-level transports,
a use that converts TCP's in-order semantics
from a convenience blessing to a performance curse.
As Internet evolution makes TCP's use as a substrate likely to grow,
we offer {\em Minion},
an architecture for backward-compatible out-of-order delivery
atop TCP and TLS\@.
Small OS API extensions
allow applications to manage TCP's send buffer
and to receive TCP segments out-of-order.
Atop these extensions,
Minion builds application-level protocols
offering true unordered datagram delivery,
within streams preserving strict wire-compatibility
with unsecured or TLS-secured TCP connections.
Minion's protocols can run on unmodified TCP stacks,
but benefit incrementally when either endpoint is upgraded,
for a backward-compatible deployment path.
Experiments suggest that Minion can noticeably improve performance
of applications such as
conferencing, virtual private networking, and web browsing,
while incurring minimal CPU or bandwidth costs.


\end{abstract}

\section{Introduction}
\label{sec:intro}

TCP~\cite{rfc793} was originally designed to offer applications
a convenient, high-level communication abstraction
with semantics emulating Unix file I/O or pipes.
As the Internet has evolved, however,
TCP's original role of offering an {\em abstraction}
has gradually been supplanted with a new role of
providing a {\em substrate}
for transport-like, application-level protocols
such as SSL/TLS~\cite{rfc5246},
	\O{}MQ~\cite{zeromq},
	SPDY~\cite{spdy}, and
	WebSockets~\cite{websocket}.
In this new {\em substrate} role,
TCP's in-order delivery offers little value
since application libraries are equally capable
of implementing convenient abstractions.
TCP's strict in-order delivery, however,
prevents applications from controlling
the {\em framing} of their communications~\cite{
	clark90architectural,ford07structured},
and incurs a ``latency tax'' on content
whose delivery must wait for the retransmission
of a single lost TCP segment.

\com{
When the network loses one data segment,
the receiving TCP must buffer and delay
all segments within at least the next round-trip time (RTT),
until the sender reacts and successfully retransmits the lost segment.
Many applications,
such as audio/video conferencing and VPN tunneling,
tolerate one packet's outright loss more gracefully
than the delay of a full RTT worth of packets,
making these applications ill-suited to TCP.
}

\xxx{We can point to section 2 and drop the following para}
Due to the difficulty of deploying new transports today~\cite{
	rosenberg08udp,ford08breaking,popa10http},
applications rarely utilize new out-of-order transports
such as SCTP~\cite{rfc4960} and DCCP~\cite{rfc4340}.
UDP~\cite{rfc768} is a popular substrate,
but is still not universally supported in the Internet,
leading even delay-sensitive applications
such as the Skype telephony system~\cite{baset06analysis}
and Microsoft's DirectAccess VPN~\cite{davies09directaccess},
to fall back on TCP despite its drawbacks.

\com{
We observe that much of TCP's latency cost is due to
send and receive buffering in the end hosts,
not in the network.
Leveraging this observation,
}

Recognizing that TCP's use as a substrate
is likely to continue and expand,
we introduce {\em Minion},
a novel architecture for efficient but backward-compatible
unordered delivery in TCP\@.
Minion consists of
{\em \utcp},
a small OS extension
adding basic unordered delivery primitives to TCP,
and two application-level protocols
implementing datagram-oriented delivery services
that function on either \utcp or unmodified TCP stacks.

\com{
{\em \utcp},
a small TCP extension
adding simple unordered delivery primitives to the OS's API,
and {\em \ucobs and \utls},
and application-level protocols
that build non-secured and secured datagram delivery substrates
atop either \utcp or unmodified TCP.
Minion is wire-compatible with TCP and TLS,
preserving compatibility with existing middleboxes
while facilitating unimpeded evolution of application-level protocols
``safely hidden'' above end-to-end TLS security.
Minion functions even if neither endpoint OS supports \utcp,
but offers incremental performance benefits
if either endpoint supports \utcp.
}

\utcp addresses delays 
caused by TCP's send and receive buffering.
On the send side,
\utcp gives the application a controlled ability
to insert data out-of-order into TCP's send queue,
allowing fresh high-priority data to bypass
previously-queued low-priority data, for example.
On the receive side,
\utcp enables the application
to receive out-of-order TCP segments immediately,
without delaying their delivery
until retransmissions fill prior holes.
Designed for simplicity and deployability,
these extensions add less than $600$ lines to Linux's TCP stack.

Minion's application-level protocols,
{\em \ucobs} and {\em \utls},
build general datagram delivery services
atop \utcp or TCP.
Key challenges these protocols address are:
(a) TCP offers no reliable out-of-band framing
to delimit datagrams in a TCP stream;
(b) \utcp cannot add out-of-band framing
without changing TCP's wire protocol;
and (c) common in-band TCP framing methods
assume in-order processing.
To make datagrams {\em self-delimiting} in a TCP stream,
\ucobs leverages
{\em Consistent Overhead Byte Stuffing} (COBS)~\cite{
	cheshire97consistent}
to encode application datagrams
with at most 0.4\% expansion,
while reserving a single byte value
to delimit encoded datagrams.

Minion adapts the stream-oriented TLS~\cite{rfc5246}
into a secure datagram delivery service atop \utcp or TCP\@.
To avoid changing the TLS wire protocol,
the \utls receiver heuristically ``guesses'' TLS record boundaries
in stream fragments received out-of-order,
then leverages TLS's cryptographic MAC
to confirm these guesses reliably.
By preserving strict wire-compatibility with TLS,
\utls enables unordered delivery within
streams indistinguishable in the network from HTTPS~\cite{rfc2818},
for example.

Experiments with a prototype on Linux
show several benefits for applications using TCP.
Minion can reduce application-perceived jitter
of Voice-over-IP (VoIP) streams atop TCP,
and increase perceptible-quality metrics~\cite{itu07wideband}.
Virtual private networks (VPNs)
that tunnel IP traffic over SSL/TLS,
such as OpenVPN~\cite{openvpn}
or DirectAccess~\cite{davies09directaccess},
can double the throughput of some tunneled TCP connections,
by employing \utcp to prioritize and expedite tunneled ACKs.
Web transports can cut the time before a page begins to appear
by up to half,
achieving the latency benefits of multistreaming transports~\cite{
	natarajan06sctp,ford07structured}
while preserving the TCP substrate.
Use of \ucobs can incur up to $5\times$ CPU load
with respect to raw TCP, due to COBS encoding,
but for secure connections,
\utls incurs less than 7\% CPU overhead (and no bandwidth overhead)
atop the baseline cost of TLS 1.1.

This paper's primary contributions are:
(a) the first wire-compatible TCP extension we are aware of
offering true out-of-order delivery;
(b) an API allowing applications to prioritize TCP's send queue;
(c) a novel use of COBS~\cite{cheshire97consistent}
for out-of-order framing atop TCP;
(d) an existence proof
that out-of-order datagram delivery is achievable
from the unmodified, stream-based TLS wire protocol;
(e) a prototype and experiments demonstrating
Minion's practicality and performance benefits.

Section~\ref{sec:motiv} motivates Minion
by discussing the evolution of TCP's role in the Internet.
Section~\ref{sec:arch} introduces Minion's high-level architecture, and
Section~\ref{sec:utcp} describes its \utcp extensions.
Section~\ref{sec:ucobs} presents \ucobs for non-secure datagram delivery,
and
Section~\ref{sec:utls} details \utls, a secure analog.
Section~\ref{sec:impl} discusses the current Minion prototype,
and
Section~\ref{sec:perf} evaluates its performance experimentally.
Section~\ref{sec:related} summarizes related work, and
Section~\ref{sec:concl} concludes.


\section{Motivating Minion}
\label{sec:motiv}

\com{ NOTES
2. Application needs and mismatches with deployed technologies
Summarize reasons why people use TCP and SSL for protocols that need
unordered services.
   2.1  Out-of-order type apps that get tunneled over TCP
        - narrow waist has moved up to TCP and SSL
        - datagram-oriented apps (voip, iSCSI, ...)
        - can we find stats on how often Skype uses TCP vs UDP?
        - Brosh et al, "The Delay-Friendliness of TCP" for making case
        that skype doesn't use UDP always
   2.2  New transports being tunneled over HTTP
        - chunking and other transport-like mechanisms
        - Why HTTP is never used atop UDP (see SST paper)
   2.3  VPNs used for tunneling apps
        - VPNs used for tunneling apps
        - Virtual/recursive networks - VPNs etc.
        - eg., Microsoft DirectAccess
   2.4  Deployment concerns for new transport services
        - Why not deployed over IP or UDP or DCCP?
        - current 3.2 (UDP)
   - point out that a lot of evidence is anecdotal and derived 
   from industry practice.
}

This section 
describes how TCP's role in the network has evolved
from a communication {\em abstraction} to a communication {\em substrate},
why its in-order delivery model makes TCP a poor substrate,
and why other OS-level transports have failed to replace TCP in this role.

\subsection{Rise of Application-Level Transports}

The transport layer's traditional role in a network stack
is to build high-level communication abstractions
convenient to applications,
atop the network layer's basic packet delivery service.
TCP's reliable, stream-oriented design~\cite{rfc793}
exemplified this principle,
by offering an inter-host communication abstraction
modeled on Unix pipes,
which were the standard {\em intra-host} communication abstraction
at the time of TCP's design.
The Unix tradition of
implementing TCP in the OS kernel
offered further convenience,
allowing much application code to ignore the difference between
an open disk file, an intra-host pipe, or an inter-host TCP socket.

\begin{figure}[tbp]
\centering
\includegraphics[width=0.47\textwidth]{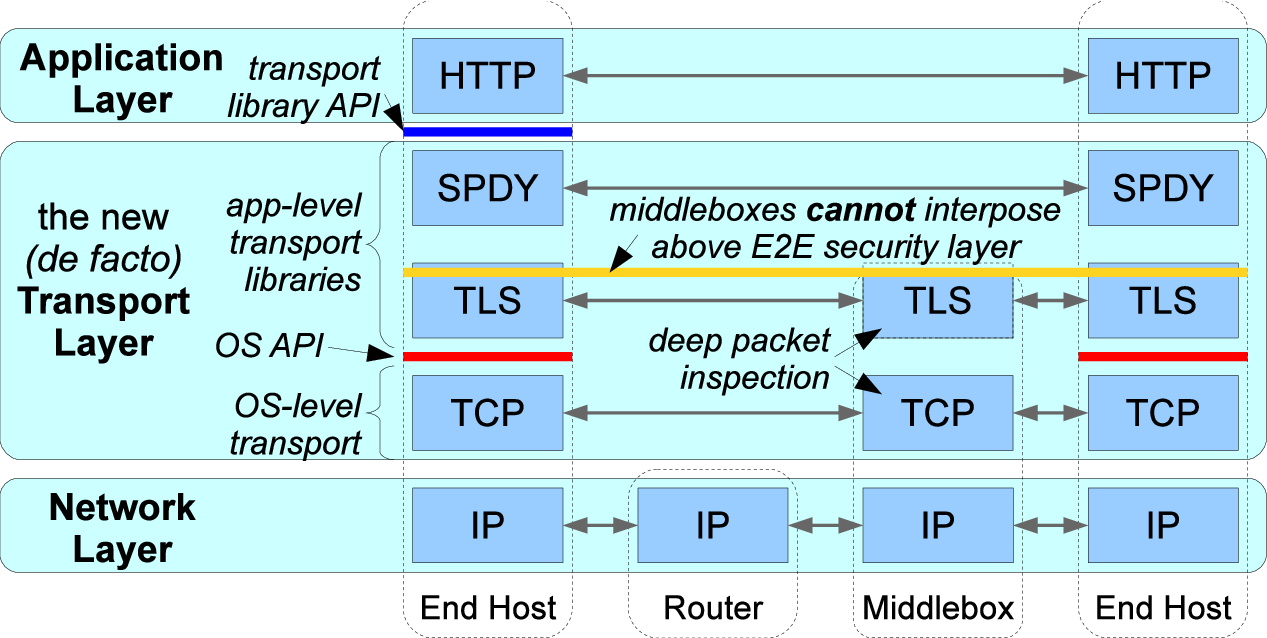}
\caption{Today's ``{\em de facto} transport layer''
	is effectively split between OS and application code.}
\label{f:layers}
\end{figure}

Instead of building directly atop
traditional OS-level transports such as TCP or UDP, however,
today's applications
frequently introduce additional transport-like
protocol layers at user-level,
typically implemented via application-linked libraries.
Examples include the ubiquitous SSL/TLS~\cite{rfc5246},
media transports such as RTP~\cite{rfc3550},
and experimental multi-streaming transports such as
SST~\cite{ford07structured}, SPDY~\cite{spdy},
and \O{}MQ~\cite{zeromq}.
Applications increasingly use HTTP or HTTPS over TCP
as a substrate~\cite{popa10http};
this is also illustrated by 
the W3C's WebSocket interface~\cite{websocket},
which offers general bidirectional communication
between browser-based applications and Web servers atop HTTP and HTTPS.

In this increasingly common design pattern,
the ``transport layer'' as a whole has in effect become
a stack of protocols straddling the OS/\linebreak[0]application boundary.
Figure~\ref{f:layers} illustrates one example stack,
representing Google's experimental Chromium browser,
which inserts SPDY for multi-streaming and TLS for security
at application level,
atop the OS-level TCP.

One can debate whether a given application-level protocol
fits some definition of ``transport'' functionality.
The important point, however,
is that today's applications no longer need, or expect,
the underlying OS to provide ``convenient'' communication abstractions:
the application simply links in libraries, frameworks,
or middleware offering the abstractions it desires.
What today's applications
need from the OS is not convenience,
but {\em an efficient substrate}
atop which application-level libraries
can build the desired abstractions.

\subsection{TCP's Latency Tax}

\com{
TCP has proven to be a popular
substrate for application-level transports.
Application protocols and libraries
such as SPDY~\cite{spdy} and
\O{}MQ~\cite{zeromq},
build concurrent, multi-streaming delivery abstractions atop TCP.
Applications increasingly use HTTP or HTTPS over TCP
as a substrate~\cite{popa10http};
the W3C's WebSocket interface~\cite{websocket},
which offers general bidirectional communication
between browser-based applications and Web servers atop HTTP and HTTPS,
illustrates the widespread demand for TCP as a substrate.
}
While TCP has proven to be a popular
substrate for application-level transports,
using TCP in this role
converts its delivery model from a blessing into a curse.
Application-level transports are just as capable as the kernel
of sequencing and reassembling packets into
a logical data unit or ``frame''~\cite{clark90architectural}.
By delaying any segment's delivery to the application
until all prior segments are received and delivered, however,
TCP imposes a ``latency tax'' on all segments
arriving within one round-trip time (RTT) after any single lost segment.

\abbr{		
This latency tax is a fundamental byproduct
}{		
\begin{figure*}[tbp]
\centering
\begin{tabular}{@{}c@{}c@{}c@{}}
\includegraphics[width=0.27\textwidth]{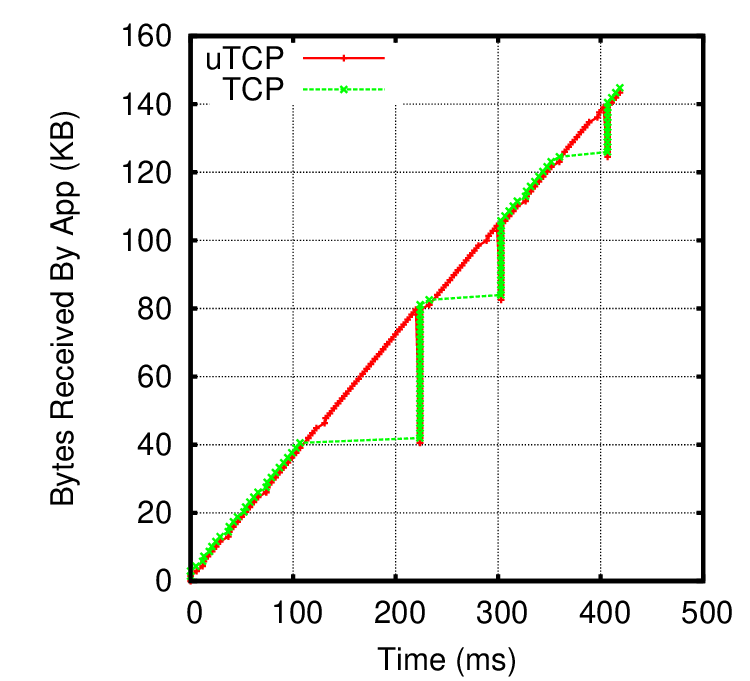} &
\includegraphics[width=0.27\textwidth]{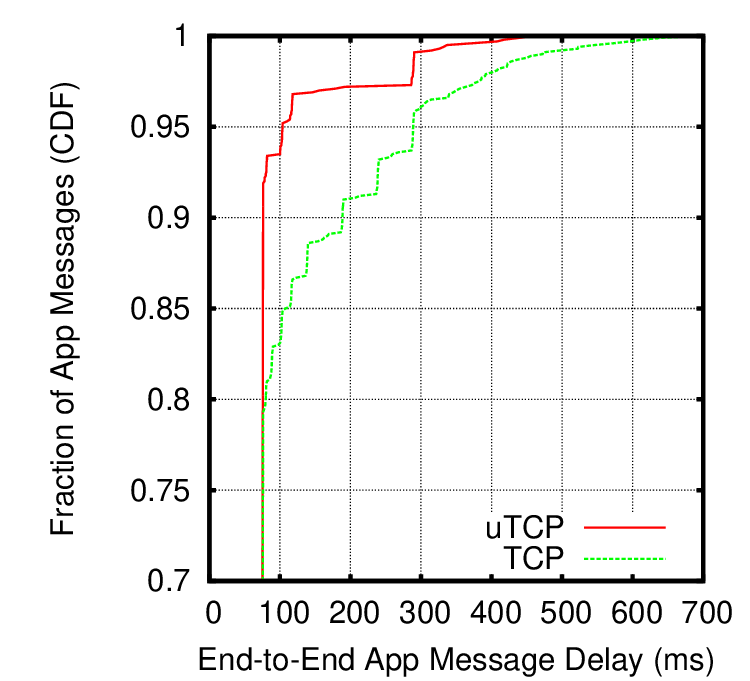} &
\includegraphics[width=0.45\textwidth]{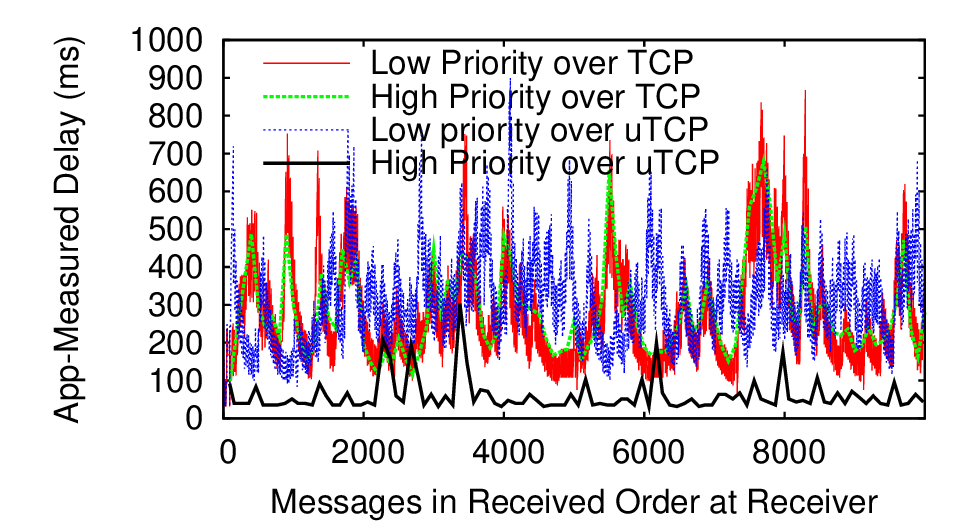} \\
{\bf (a)} Application-observed
		& {\bf (b)} CDF of app-observed
				& {\bf (c)} Latency improvements achievable \\
progress over time	
		& message delivery time
				& with send buffer prioritization	\\
\end{tabular}

\caption{
Graphs illustrating some potential benefits from unordered delivery in TCP.
}
\label{f:works}
\end{figure*}

\com{
\begin{figure}[tbp]
\centering
\includegraphics[width=0.4\textwidth]{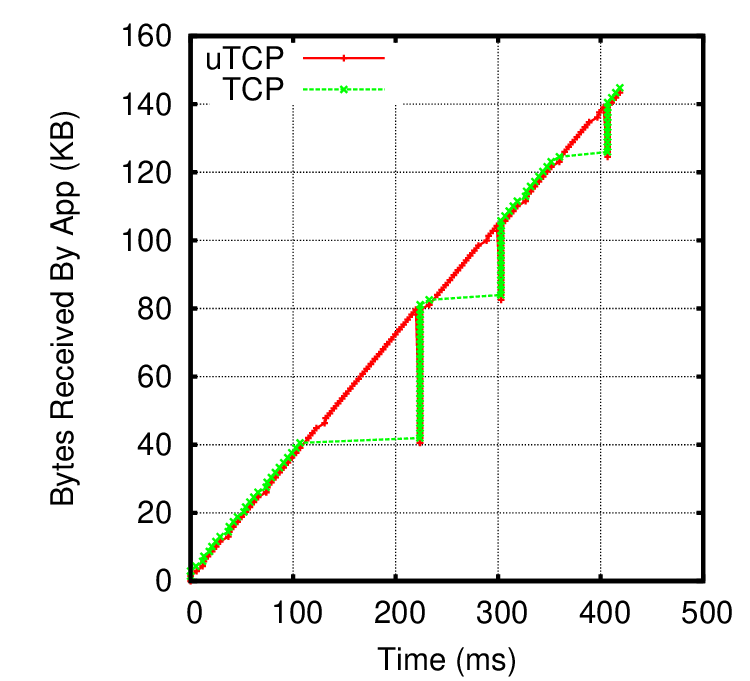}
\caption{
Bytes received by an application over time,
in-order atop TCP versus out-of-order atop \utcp.}
\label{f:motiv-progress}
\end{figure}

\begin{figure}[tbp]
\centering
\includegraphics[width=0.4\textwidth]{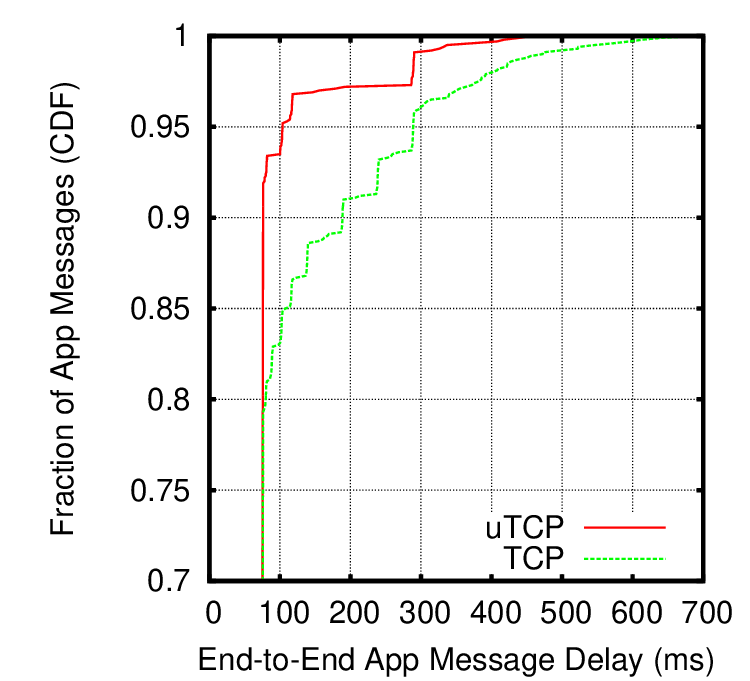}
\caption{
CDF of application-observed message delivery latency
via in-order TCP versus out-of-order \utcp.}
\label{f:motiv-latency}
\end{figure}
}

Figure~\ref{f:works}(a) illustrates TCP's latency tax,
showing cumulative bytes delivered over time
during a simple bulk transfer
atop TCP versus the unordered \utcp substrate we introduce later.
The transfer runs on a network path with $60$ms RTT,
with an unrealistically high $3$\% loss
to make several losses appear in the short duration.
From the application's perspective,
all receive progress stops for one or more
round-trip times after any segment's loss,
while \utcp continues delivery without interruption.

Figure~\ref{f:works}(b)
shows a CDF of application-observed delivery latencies
in an experiment suggestive of streaming video,
where the sender transmits $1448$-byte records
at fixed $20$ms intervals,
over a path $100$ms RTT and $2$\% random loss.
Atop TCP, over $10$\% of records
are delayed by at least one full RTT,
waiting in TCP's receive queue
for a prior segment's retransmission.
Unordered delivery atop \utcp significantly delays
fewer than $3$\% of records.

Figure~\ref{f:works}(c),
showing application-observed message delay over message number,
illustrates how \utcp's send-side prioritization
affects a series of high-priority application messages
interspersed with constant low-priority traffic
atop a network-limited TCP connection.
In this simple experiment,
conducted over a network path
with a $60$ms RTT and $0.5$\% loss,
unmodified TCP delays high- and low-priority traffic equally,
but \utcp enables high-priority messages
to short-cut the send queue
and achieve lower latency.

These experiments
merely illustrate TCP's ``latency tax'';
we defer to Section~\ref{sec:perf} for
further analysis of TCP's latency effects under more realistic conditions.
For now,
the important point is that
this latency tax is a fundamental byproduct
}
of TCP's in-order delivery model,
and is irreducible,
in that an application-level transport
cannot ``claw back''
the time a potentially useful segment
has wasted in TCP's buffers.
The best the application can do is
simply to {\em expect} higher latencies to be common.
A conferencing application can use a longer jitter buffer,
for example,
at the cost of increasing user-perceptible lag.
Network hardware advances are unlikely to address this issue, 
since TCP's latency tax depends on RTT,
which is lower-bounded by the speed of light
for long-distance communications.

\subsection{Alternative OS-level Transports}
\label{sec:motiv-alts}

All standardized OS-level transports since TCP, including
	UDP~\cite{rfc768}, RDP~\cite{rfc908},
	DCCP~\cite{rfc4340}, and SCTP~\cite{rfc4960},
support out-of-order delivery.
The Internet's evolution has created strong barriers
against the widespread deployment of new transports
other than the original TCP and UDP, however.
These barriers are detailed elsewhere~\cite{
	rosenberg08udp,ford08breaking,popa10http},
but we summarize two key issues here.

\abbr{ 
First, 
adding or enhancing a ``native'' transport built atop IP
involves modifying popular OSes,
effectively increasing the bar for widespread deployment
and making it 
more difficult to evolve transport functionality
below the red line representing the OS API
in Figure~\ref{f:layers}.
Second, the Internet's original ``dumb network'' design,
in which routers that ``see'' only up to the IP layer,
has evolved into a ``smart network''
in which pervasive middleboxes
perform deep packet inspection and interposition
in transport and higher layers.
Firewalls tend to block ``anything unfamiliar'' for security reasons,
and Network Address Translators (NATs)
rewrite the port number in the  transport header,
making both incapable of allowing traffic from a new transport
without explicit support for that transport.
Any packet content
not protected by end-to-end security such as TLS---%
the yellow line in Figure~\ref{f:layers}---%
has become ``fair game''
for middleboxes to inspect and interpose on~\abcite{
	reis08detecting}{vratonjic10integrity},
making it more difficult to evolve transport functionality
anywhere below that line.

}{ 
First, since mainstream operating systems
offer unprivileged network access
only above the transport layer,
adding or enhancing a ``native'' transport built atop IP
involves modifying the OS,
creating a chicken-and-egg problem for deployment.
OS vendors are reluctant to add or modify transports
until they perceive widespread demand from applications,
but applications are reluctant to rely on new transports
without widespread OS support.
In effect, it is much more difficult to evolve transport functionality
below the red line representing the OS API
in Figure~\ref{f:layers}.

Second, the Internet's original ``dumb network'' design,
in which routers that ``see'' only up to the IP layer,
has evolved into a ``smart network''
in which pervasive middleboxes
perform deep packet inspection and interposition
in transport and higher layers.
Firewalls tend to block ``anything unfamiliar'' for security reasons,
often including new transport or application protocols
as collateral damage.
Network address translators
need to rewrite the port numbers in transport layer headers,
making them unable to carry traffic for a new or unknown transport
without explicit support for that transport.
Any packet content
not protected by an end-to-end security layer such as TLS---%
the yellow line in Figure~\ref{f:layers}---%
has in effect become ``fair game''
for middleboxes to inspect and interpose on~\abcite{
	reis08detecting}{vratonjic10integrity},
making it more difficult to evolve transport functionality
anywhere below that line.
}

\subsection{Why Not UDP?}

As the only widely-supported transport
with out-of-order delivery,
UDP offers a natural
substrate for application-level transports.
Even applications otherwise well-suited to UDP's delivery model
often favor TCP as a substrate, however.
A recent study found over 70\% of streaming media using TCP~\cite{
	guo06delving},
and even latency-sensitive conferencing applications
such as Skype
often use TCP~\cite{baset06analysis}.

Network middleboxes support UDP widely but not {\em universally}.
For this reason,
latency-sensitive applications
seeking maximal connectivity ``in the wild''
often fall back to TCP when UDP connectivity fails.
Skype~\cite{baset06analysis}
and Microsoft's Direct\-Access VPN~\cite{davies09directaccess},
for example,
support UDP but can masquerade
as HTTP or HTTPS streams atop TCP when required for connectivity.

TCP can offer performance advantages over UDP as well.
For applications requiring congestion control,
an OS-level implementation in TCP
may be more timing-accurate
than an application-level implementation in a UDP-based protocol,
because the OS kernel
can avoid the timing artifacts of 
system calls and process scheduling~\cite{zec02estimating}.
Hardware TCP offload engines
can optimize common-case efficiency
in end hosts~\cite{mogul03tcp},
and performance enhancing proxies
can optimize TCP throughput across diverse networks~\cite{
	rfc3234,cisco-rbscp}.
Since middleboxes can track TCP's state machine,
they impose much longer idle timeouts on open TCP connections---%
nominally two hours~\cite{rfc5382}---%
whereas UDP-based applications must send keepalives
every two minutes to keep an idle connection open~\cite{rfc4787},
draining power on mobile devices.

For applications,
TCP versus UDP represents an ``all-or-nothing'' choice
on the spectrum of services applications need.
Applications desiring some but not all of TCP's services,
such as congestion control but unordered delivery,
must reimplement and tune all other services atop UDP
or suffer TCP's performance penalties.

Without dismissing UDP's usefulness 
as a truly ``least-common-denominator'' substrate,
we believe the factors above
suggest that TCP will also remain a popular substrate---%
even for latency-sensitive applications
that can benefit from out-of-order delivery---%
and that a deployable, backward-compatible workaround
to TCP's latency tax
can significantly benefit such applications.


\section{Minion Architecture Overview}
\label{sec:arch}

\begin{figure}[tbp]
\centering
\includegraphics[width=0.47\textwidth]{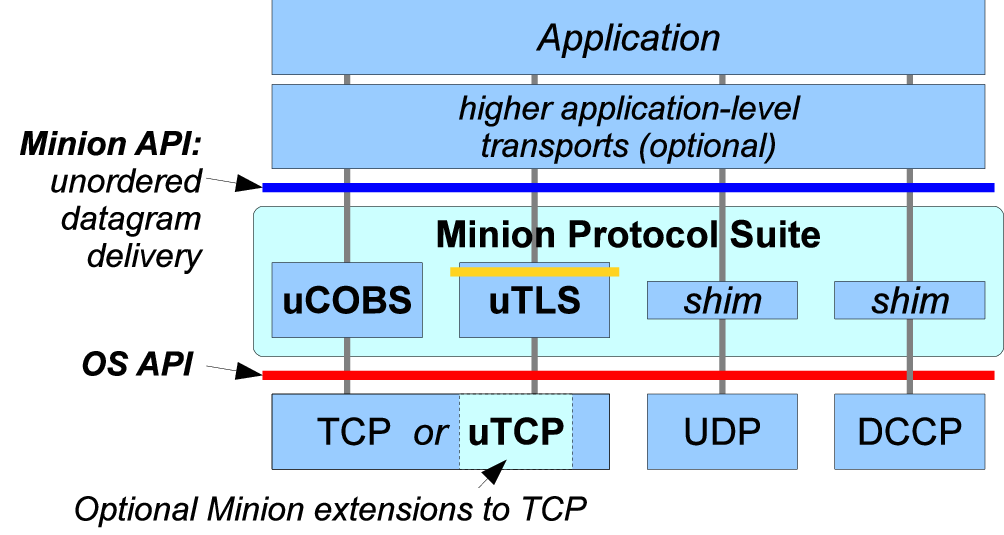}
\caption{Minion architecture}
\label{f:arch}
\end{figure}

Minion is an architecture and protocol suite
designed to meet the needs of today's applications
for efficient unordered delivery built atop either TCP or UDP.
Minion itself offers no high-level abstractions:
its goal is to serve
applications and higher application-level transports,
by acting as a ``packhorse''
carrying raw datagrams as reliably and efficiently as possible
across today's diverse and change-averse Internet.

\subsection{All-Terrain Unordered Delivery}

Figure~\ref{f:arch} illustrates Minion's architecture.
Applications and higher application-level transports
link in and use Minion in the same way as they already use
existing application-level transports
such as DTLS~\cite{rfc4347},
the datagram-oriented analog of SSL/TLS~\cite{rfc5246}.
In contrast with DTLS's goal of layering security
atop datagram transports such as UDP or DCCP,
Minion's goal is to offer efficient datagram delivery
atop {\em any} available OS-level substrate, including TCP.

While many protocols embed datagrams
or application-level frames
into TCP streams using delimiting schemes,
to our knowledge Minion is the first application-level transport that,
under suitable conditions,
offers true unordered delivery atop TCP.
Minion effectively offers relief from TCP's latency tax:
the loss of one TCP segment in the network
no longer prevents datagrams embedded in subsequent TCP segments
from being delivered promptly to the application.

\subsection{Minion Architecture Components}

Minion consists of
several application-level transport protocols,
together with a set of optional enhancements
to end hosts' OS-level TCP implementations.

Minion's enhanced OS-level TCP stack,
which we call {\em \utcp} (``unordered TCP''),
includes sender- and receiver-side
API features supporting unordered delivery and prioritization,
detailed in Section~\ref{sec:utcp}.
These enhancements affect only the OS API
through which application-level transports such as Minion
interact with the TCP stack,
and make {\em no} changes to TCP's wire protocol.

\abbr{	
Minion's application-level protocol suite currently consists
of \ucobs,
which implements unordered datagram delivery
atop unmodified TCP or \utcp streams
using COBS encoding~\cite{cheshire97consistent}
as described in Section~\ref{sec:ucobs};
and \utls,
which adapts the traditionally stream-oriented TLS~\cite{rfc5246}
into a secure unordered datagram delivery service
atop TCP or \utcp.
Minion also adds trivial shim layers
atop OS-level datagram transports, such as UDP and DCCP,
to give applications a consistent API
for unordered delivery across multiple OS-level transports.
}{	
Minion's application-level protocol suite currently consists
of the following main components:
\begin{itemize}
\item	{\bf \ucobs} is a protocol
	that implements a minimal unordered datagram delivery service
	atop either unmodified TCP or \utcp,
	using COBS encoding~\cite{cheshire97consistent}
	to facilitate out-of-order datagram delimiting
	and prioritized delivery,
	as described later in Section~\ref{sec:ucobs}.
\item	{\bf \utls} is a modification of
	the traditionally stream-oriented TLS~\cite{rfc5246},
	offering a secure, unordered datagram delivery service
	atop TCP or \utcp.
	The wire-encoding of \utls streams
	is designed to be indistinguishable in the network
	from conventional, encrypted TLS-over-TCP streams (e.g., HTTPS),
	offering a maximally conservative design point
	that makes no network-visible changes
	``below the yellow line''
	in Figure~\ref{f:layers}.
	Section~\ref{sec:utls} describes \utls.
\item	Minion adds shim layers
	atop OS-level datagram transports, such as UDP and DCCP,
	to offer applications a consistent API
	for unordered delivery across multiple OS-level transports.
	Since these shims are merely wrappers
	for OS transports already offering unordered delivery,
	this paper does not discuss them in detail.
\end{itemize}
}

Minion currently leaves to the application
the decision of {\em which} protocol to use for a given connection:
e.g., \ucobs or \utls atop TCP/\utcp,
or OS-level UDP or DCCP via Minion's shims.
We are developing an experimental {\em negotiation protocol}
to explore the protocol configuration space dynamically,
optimizing protocol selection and configuration
for the application's needs and the network's 
constraints~\cite{ford09efficient},
but we defer this enhancement to future work.
Many applications already incorporate simple negotiation schemes,
however---%
e.g., attempting a UDP connection first
and falling back to TCP if that fails---%
and adapting these mechanisms
to engage Minion's protocols according to
application-defined preferences and decision criteria
should be straightforward.

\subsection{Compatibility and Deployability}

Minion addresses the key barriers to transport evolution,
outlined in Section~\ref{sec:motiv-alts},
by creating a backward-compatible, incrementally deployable substrate
for new application-layer transports desiring unordered delivery.
Minion's deployability rests on the fact that it can,
when necessary,
avoid relying on changes either
``below the red line'' in the end hosts
(the OS API in Figure~\ref{f:layers}),
or
``below the yellow line'' in the network
(the end-to-end security layer in Figure~\ref{f:layers}).

While Minion's \ucobs and \utls protocols
offer maximum performance benefits from out-of-order delivery
when both endpoints include OS support for Minion's \utcp enhancements,
\ucobs and \utls still function and interoperate correctly
even if neither endpoint supports \utcp,
and the application need not know or care
whether the underlying OS supports \utcp.
If only one endpoint OS supports \utcp,
Minion still offers incremental performance benefits,
since \utcp's sender-side and receiver-side enhancements are independent.
A \ucobs or \utls connection atop a mixed TCP/\linebreak[0]\utcp endpoint-pair
benefits from \utcp's sender-side enhancements
for datagrams sent by the \utcp endpoint,
and the connection benefits from \utcp's receiver-side enhancements
for datagrams arriving at the \utcp host.

\abbr{}{
\begin{figure}[htbp]
\centering
\includegraphics[width=0.47\textwidth]{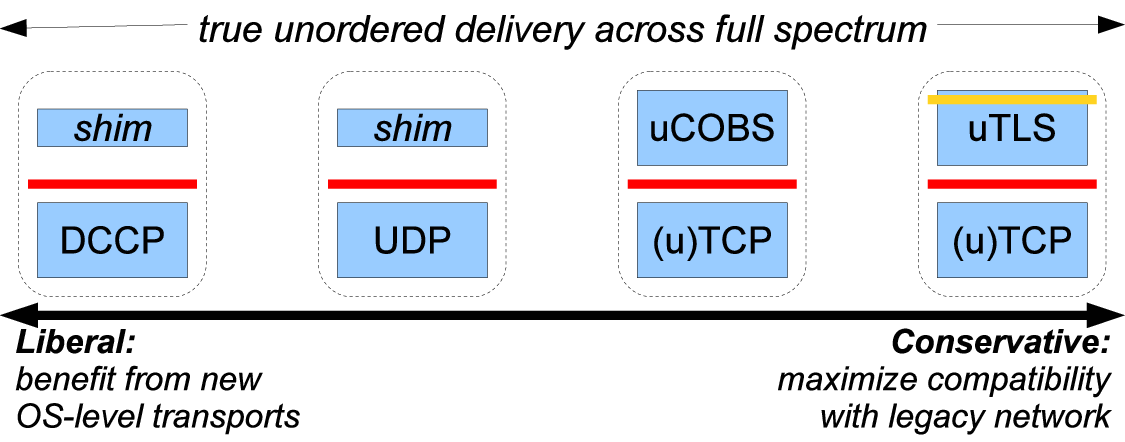}
\caption{Continuum of Minion configuration tradeoffs}
\label{f:continuum}
\end{figure}
}

Addressing the challenge of network-compatibility
with middleboxes that filter new OS-level transports
and sometimes UDP,
Minion offers application-level transports
a continuum of substrates
representing different tradeoffs
between suitability to the application's needs
and compatibility with the network.
\abbr{
An application can use unordered OS-level transports
such as UDP, DCCP~\cite{rfc4340}, or SCTP~\cite{rfc4960},
for paths on which they operate,
but Minion offers an unordered delivery alternative
usable even when TCP is the only viable choice.
}{
Figure~\ref{f:continuum} illustrates this continuum
of Minion configurations.
At the ``liberal'' end,
an application can still use whatever OS-level transport
is most suited to its purposes,
such as UDP, DCCP~\cite{rfc4340}, or SCTP~\cite{rfc4960},
for paths on which these unordered OS-level transports operate.

On the ``moderate conservative'' side,
\ucobs offers an unordered substrate
wire-compatible with TCP up through the OS-level transport layer.
At the ``extreme conservative'' end,
\utls offers an unordered substrate
wire-compatible not only with TCP,
but with the ubiquitous TLS-over-TCP streams
on which HTTPS (and hence Web security and E-commerce) are based,
likely to operate in almost any network environment
purporting to offer ``Internet access.''
}

A final issue is compatibility with existing applications.
Since most of Minion operates at application-level,
applications must be changed to use the Minion API.
A pair of application endpoints may also need to negotiate
whether to use Minion,
or to run directly atop OS-level transports
for compatibility with earlier versions of the application.
This challenge is comparable to
the cost of adding TLS or DTLS support to an application,
and the popularity of application-level transports such as TLS
suggests that these costs are surmountable.
Minion's application-level functionality
might eventually be merged into existing or future
application-level transports and communication frameworks,
making its benefits available with few or no application changes.

\section{\utcp: Unordered TCP}
\label{sec:design}
\label{sec:utcp}

Minion enhances the OS's TCP stack
with API enhancements supporting unordered delivery
in both TCP's send and receive paths,
enabling applications to reduce transmission latency
at both the sender- and receiver-side end hosts
when both endpoints support \utcp.
Since \utcp makes no changes to TCP's wire protocol,
two endpoints need not ``agree'' on whether to use \utcp:
one endpoint gains latency benefits from \utcp
even if the other endpoint does not support it.
Further, an OS may choose independently
whether to support the sender- and receiver-side enhancements,
and when available, applications can activate them independently.

In this spirit of Section~\ref{sec:motiv},
\utcp does {\em not} seek to offer
``convenient'' or ``clean'' unordered delivery abstractions
directly at the OS API.
Instead, \utcp's design is motivated by the goals of
maintaining exact compatibility
with TCP's existing wire-visible protocol and behavior,
and facilitating deployability
by minimizing the extent and complexity
of changes to the OS's TCP stack.
The design presented here is only one of many viable approaches,
with different tradeoffs,
to supporting unordered delivery in TCP.
Section~\ref{sec:utcp-alts} briefly outlines a few such alternatives.

\abbr{}{
\utcp focuses on reducing latency only at the end hosts.
There are many ways to reduce TCP latency in the network,
such as using delay-based or explicit congestion control~\cite{
	brakmo95tcp,katabi02internet}.
These valuable efforts are independent of and complementary
to Minion's goals and are not addressed here.
}

We describe \utcp's API enhancements
in terms of the BSD sockets API,
although \utcp's design contains nothing inherently specific
to this API.


\subsection{Receiver-Side Enhancements}
\label{sec:utcp-recv}

\begin{figure*}[htb]
\centering
\includegraphics[width=0.99\textwidth]{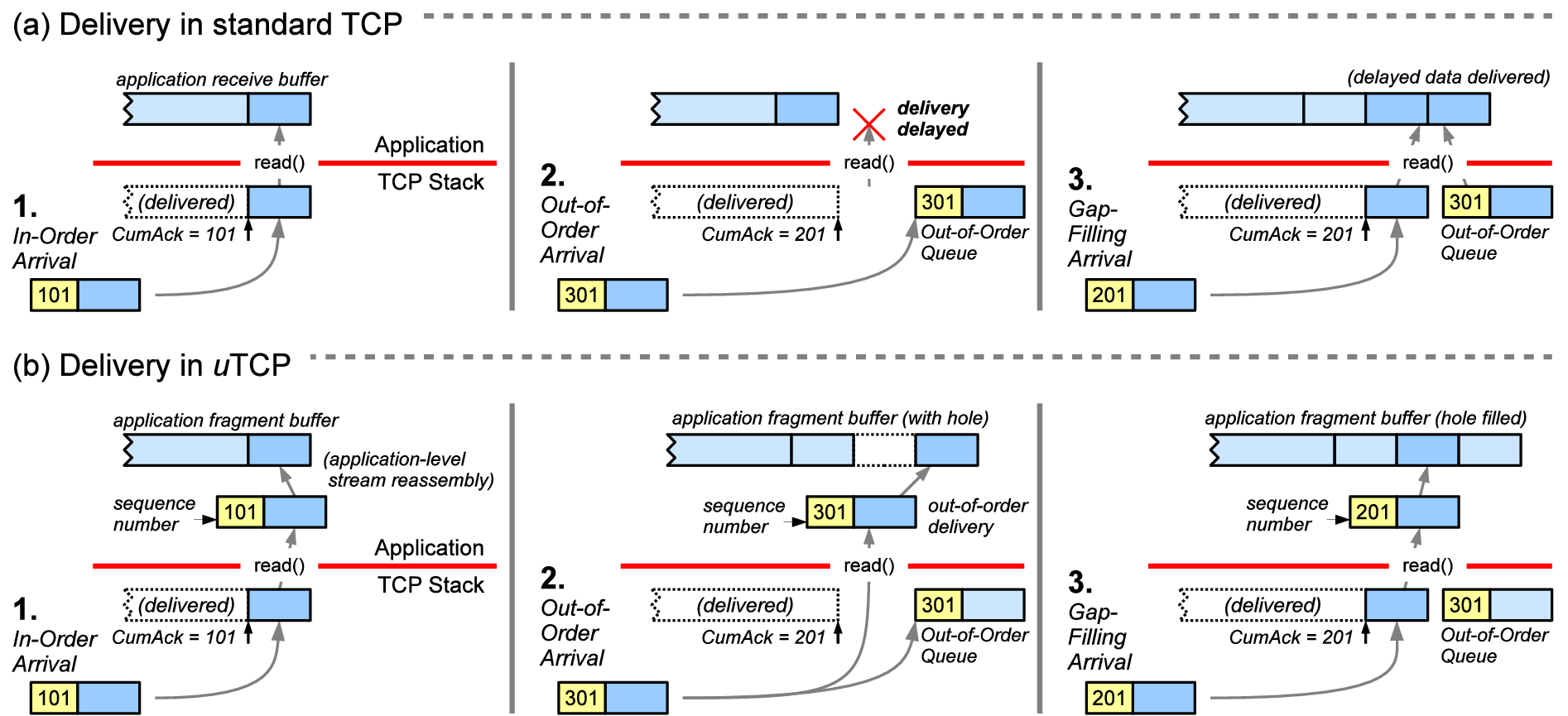}
\caption{
Delivery behavior of (a) standard TCP, and (b) \utcp,
upon receipt of in-order and out-of-order segments.
}
\label{f:delivery}
\end{figure*}

\utcp adds one new socket option
affecting TCP's receive path,
enabling applications to request immediate delivery
of TCP segments received out of order.
An application opens a TCP stream the usual way,
via \verb|connect()| or \verb|accept()|,
and may use this stream for conventional in-order communication
before enabling \utcp.
Once the application is ready to receive out-of-order data,
it enables the new option \verb|SO_UNORDERED|
via \verb|setsockopt()|,
which changes TCP's receive-side behavior in two ways.

First, whereas a conventional TCP stack
delivers received data to the application
only when prior gaps in the TCP sequence space are filled,
the \utcp receiver
makes data segments available to the application immediately upon receipt,
skipping TCP's usual reordering queue.
The application obtains this data via \verb|read()| as usual,
but the first data byte returned by a \verb|read()| call
may no longer be the one
logically following the last byte returned
by the prior \verb|read()| call,
in the byte stream transmitted by the sender.
The data the \utcp stack delivers to the application
in successive \verb|read()| calls
may skip forward and backward in the transmitted byte stream,
and \utcp may even deliver portions
of the transmitted stream multiple times.
\utcp guarantees only that the data returned by one \verb|read()| call
corresponds to {\em some} contiguous sequence of bytes
in the sender's transmitted stream,
and that barring connection failure,
\utcp will {\em eventually} deliver
every byte of the transmitted stream at least once.

Second,
when servicing an application's \verb|read()| call,
the \utcp receiver prepends a short header to the returned data,
indicating the logical offset of the first returned byte
in the sender's original byte stream.
The \utcp stack computes this logical offset
simply by subtracting the
Initial Sequence Number (ISN) of the received stream
from the TCP sequence number of the segment being delivered.
Using this metadata, the application can piece together
data segments from successive \verb|read()| calls
into longer contiguous {\em fragments} of the transmitted byte stream.

Figure~\ref{f:delivery} illustrates \utcp's receive-side behavior,
in a simple scenario
where three TCP segments arrive in succession:
first an in-order segment, then an out-of-order segment,
and finally a segment filling the gap between the first two.
With \utcp,
the application receives each segment as soon as it arrives,
along with the sequence number information it needs to reconstruct
a complete internal view of
whichever fragments of the TCP stream have arrived.

The \utcp receiver retains in its receive buffer
the TCP headers of segments received and delivered out-of-order,
until its cumulative acknowledgment point
moves past these segments,
and generates acknowledgments and selective acknowledgments (SACKs)
exactly as TCP normally would.
The \utcp receiver does not increase its advertised receive window
when it delivers data to the application out-of-order,
so the advertised window tracks the cumulative in-order delivery point
exactly as in TCP.
In this fashion,
\utcp maintains wire-visible behavior identical to TCP
while delivering segments to the application out-of-order.

The \utcp receive path assumes the sender may be an unmodified TCP,
and TCP's stream-oriented semantics allow the sending TCP
to segment the sending application's stream at arbitrary points---%
independent of the boundaries of
the sending application's \verb|write()| calls, for example.
Further, network middleboxes may
silently {\em re-segment} TCP streams,
making segment boundaries observed at the receiver
differ from the sender's original transmissions~\cite{
	honda11possible}.
An application using \utcp,
therefore,
must not assume anything about received segment boundaries.
This is a key technical challenge to using \utcp reliably,
and addressing this challenge is one function
of \ucobs and \utls, described later.

\subsection{Sender-Side Enhancements}
\label{sec:utcp-send}

While \utcp's receiver-side enhancements
address the ``latency tax''
on segments waiting in TCP's reordering buffer,
TCP's sender-side queue can also introduce latency,
as segments the application has already written to a TCP socket---%
and hence ``committed'' to the network---%
wait until TCP's flow and congestion control
allow their transmission.
Many applications can benefit from the ability to ``late-bind''
their decision on {\em what} to send until the last possible moment.
An application-level multi-streaming transport with prioritization,
such as SST~\cite{ford07structured} or SPDY~\cite{spdy}, would prefer high-priority packets not to get ``stuck'' behind low-priority packets in TCP's send queue.
In applications such as games and remote-access protocols,
where the receiver typically desires {\em only the freshest}
in a stream of real-time status updates,
the sender would prefer that new updates ``squash'' any prior updates
still in TCP's send queue and not yet transmitted.

The Congestion Manager architecture~\cite{balakrishnan99integrated}
addressed this desire
to ``late-bind'' the application's transmission decisions,
by introducing an upcall-based API
in which the OS performs no send buffering,
but instead signals the application
whenever the application is permitted to send.
Upcalls represent a major change to conventional sockets APIs,
however,
and introduce issues such as how the OS should handle an application
that fails to service an upcall promptly,
leaving its allocated transmission time-slot unfilled
yet unavailable to competing applications waiting to send.

In the spirit of maximizing deployability,
\utcp adopts a more limited but less invasive design,
by retaining TCP's send buffer
but giving applications some control over it.
After enabling \utcp's new socket option \verb|SO_UNORDEREDSEND|,
the OS expects any subsequent \verb|write()| to that socket
to include a short header,
containing metadata that \utcp reads and strips
before placing the remaining data on TCP's send buffer.
The \utcp header contains an integer {\em tag}
and a set of optional flags
controlling \utcp's send-side behavior.

By default,
\utcp interprets tags as priority levels.
Instead of unconditionally placing the newly-written data
at the tail of the send queue as TCP normally would,
\utcp{} {\em inserts} the newly-written data into the send queue
just {\em before} any lower-priority data
in the send queue and not yet transmitted.
The application thus avoids higher-priority packets
being delayed by lower-priority packets enqueued earlier,
while the OS avoids the complexity and security challenges
of an upcall API.

For strict TCP wire-compatibility,
\utcp never inserts new data into the send queue
ahead of any previously-written data that has ever been transmitted
in whole or in part:
e.g., ahead of data from a prior application write
already partly transmitted and awaiting acknowledgment.
If an application writes a large low-priority buffer,
then writes higher-priority data
after transmission of the low-priority data has begun,
\utcp inserts the high-priority data after
the entire low-priority write and never in the middle.
This constraint enables the application to control
the boundaries on which send buffer reordering is permitted,
independent of the current MTU and TCP segmentation behavior.

A simple \utcp refinement,
which we intend to explore in future work,
is to include a {\em squash} flag
in the metadata header the application prepends to each write.
If set,
while inserting the newly-written data
in tag-priority order,
\utcp would also remove and discard any data
previously written with exactly the same tag,
that has not yet been transmitted in whole or in part.
This refinement would enable update-oriented applications
such as games to avoid the bandwidth cost
transmitting old updates superseded by newer data.

\subsection{Design Alternatives}
\label{sec:utcp-alts}


\utcp pursues a conservative point in a large design space,
and many alternatives present interesting tradeoffs.
\abbr{	
Some alternatives include:
disabling TCP congestion control at the sender;
assigning TCP sequence numbers at application write time
instead of the time a segment is first transmitted;
sending new data in retransmitted segments;
modifying the receiver to acknowledge unreceived sequence space gaps
for unreliable service;
increasing the receive window to account for out-of-order segments;
and delivering data to the application exactly-once instead of at-least-once.
For space reasons we discuss these tradeoffs
in more detail elsewhere~\cite{iyengar12minion-full}.
A common theme, however, is that most of these design alternatives
change TCP's behavior in wire-visible ways,
which can trigger various unpredictable middlebox behaviors~\cite{
	honda11possible},
making connectivity less reliable.

}{	
We briefly examine a few such alternatives.

\paragraph{Disable TCP congestion control}

By default,
\utcp makes no changes to TCP congestion control,
making \utcp's unordered delivery model
more directly comparable to DCCP~\cite{rfc4340}
than to UDP~\cite{rfc768}.
Some applications may prefer a UDP-like substrate
without congestion control:
e.g., media applications
that adapt to congestion by transitioning
between a set of fixed bit-rates.
Purely for experimentation,
we extended TCP's API
to enable the application to disable congestion control at the sender
per-connection---%
as Linux has already supports
per-interface~\cite{NoQ}---%
offering applications more UDP-like behavior.
Such extensions are independent and orthogonal to
the unordered delivery extensions we focus on here, however,
and we take no position on the policy question
of whether the OS {\em should} allow applications
to disable congestion control.

\paragraph{Assign Sequence Numbers at Write Time}
All application-directed reordering of the send buffer
currently occurs {\em before} \utcp assigns sequence numbers to segments.
Thus, TCP sequence numbers appearing on the wire
still reflect transmission order from the sending {\em host},
exactly preserving TCP's wire-visible behavior.
An alternative would be for \utcp to assign TCP sequence numbers
immediately as the application {\em writes} to the send buffer,
preserving these sequence numbers through send buffer reordering.
This alternative could simplify some applications,
as discussed later in Section~\ref{sec:utls-send}.
Send buffer reordering would instead mimic reordering in the network,
however, which could violate the common assumption
that reordering in the network is rare.
Frequent reordering due to send-side prioritization
could cause TCP segments to hit ``slow paths''
in middleboxes and the receiving TCP stack more frequently,
for example,
and could interfere with the sender's congestion control
by creating the illusion of false loss events,
unless the congestion control scheme is adjusted to compensate.
Assigning sequence numbers immediately upon write
would also preclude the {\em squash} flag refinement above.

\paragraph{Disable retransmission of old TCP segments}
The \utcp receiver
could always proactively move its cumulative acknowledgment point
past {\em all} received segments,
including those received out-of-order,
effectively pretending that lost segments were received
and tricking the sender into not retransmitting them.
This change would enable applications
to avoid the bandwidth cost
of retransmitting obsolete data.
As a side-effect,
this change would interfere with the sender's congestion control
by effectively hiding all ``loss events''
other than retransmission timeouts,
which might or might not be desirable.
This change would also affect wire-visible behavior, however:
middleboxes could observe the receiver acknowledging
TCP sequence number ``holes''
whose data was dropped upstream from the middlebox,
resulting in unpredictable middlebox behavior~\cite{honda11possible}.

\com{
Second,
like TCP,
\utcp still retransmits {\em all} application data
until the receiver acknowledges it
or the connection fails.
Unlike UDP or DCCP applications,
a \utcp application cannot avoid the bandwidth cost
of retransmitting application data
that may be obsolete by the time it finally arrives,
such as a late video frame.
Since TCP's congestion control is well-known to perform poorly
at high packet loss rates, however,
we expect the bandwidth cost of these unnecessary retransmissions
to be small whenever \utcp is operating over a network path
on which TCP would perform reasonably.
The next section discusses two design alternatives
that could eliminate these unnecessary retransmissions,
but at considerable costs in network compatibility.
}

\paragraph{Send new data in retransmission segments}

\utcp could send new application data,
instead of the data originally sent,
in retransmissions of unacknowledged sequence ranges.
This change would eliminate
the cost of retransmitting obsolete data,
and the need for the sending TCP
to buffer transmitted data until acknowledgment.
Network delays, reordering, and segmentation
could result in the receiver seeing a mix of ``new'' and ``old'' data
spliced at arbitrary boundaries,
however,
and inconsistent retransmission can trigger
unpredictable middlebox behavior~\cite{
	honda11possible}.

\paragraph{Increase the receive window on out-of-order delivery}

The \utcp receiver technically need not buffer out-of-order data
once it has been delivered to the application---%
only the sequence ranges it needs to send correct ACKs and SACKs.
Since the TCP receive window announcement
traditionally reflects the amount of buffer space available at the receiver,
the \utcp receiver could increase its receive window announcement
when it delivers out-of-order data,
as TCP does when it delivers in-order data.
This change may introduce a denial-of-service attack vulnerability, however,
where a sender keeps sending data out-of-order to a \utcp stack indefinitely
without ever going back to ``fill the gaps''
and advance the cumulative acknowledgment point,
leading to unbounded state in the \utcp stack and possibly in the application.
In \utcp's more conservative design,
the receive window imposes a limit on the number of out-of-order bytes
outstanding before the sender must retransmit lost data
and move the cumulative-acknowledgment point forward.

\paragraph{Offer the application exactly-once data delivery}

\utcp could guarantee the delivery of
a given byte of the transmitted stream
{\em exactly} once, rather than at least once,
by ``pruning'' data already received
from subsequent out-of-order and in-order deliveries.
Unlike the above design changes,
this one would not affect network-visible behavior
and may be worthwhile.
But it would increase the complexity
of \utcp's modifications to the TCP stack,
for dubious benefit to the application.
A \utcp application must in any case contain the state and logic necessary
to assemble segments received out-of-order into larger fragments
and scan these fragments for useful application records.
On top of this,
we find the incremental complexity cost
of detecting and ignoring duplicate data
at application level to be small.

}



\section{\ucobs: Simple Datagrams on TCP}
\label{sec:ucobs}

Since \utcp's design attempts to minimize OS changes,
its unordered delivery primitives do not directly offer applications
a convenient, general-purpose datagram substrate.
Minion's \ucobs protocol bridges this semantic gap,
building atop \utcp (or standard TCP)
a lightweight datagram delivery service
comparable to UDP or DCCP.
This first section first introduces the challenge of delimiting datagrams,
then presents \ucobs' solution and discusses alternatives.

\subsection{Self-Delimiting Datagrams for \utcp}

Applications built on datagram substrates such as UDP
generally assume the underlying layer preserves datagram boundaries.
If the network fragments a large UDP datagram,
the receiving host reassembles it before delivery to the application,
and a correct UDP never merges multiple datagrams, or datagram fragments,
into one delivery to the receiving application.
TCP's stream-oriented semantics do not preserve
any application-relevant frame boundaries within a stream, however.
Both the TCP sender and network middleboxes can and do
coalesce TCP segments or re-segment TCP streams
in unpredictable ways~\cite{honda11possible}.
Conventional TCP applications,
which send and receive TCP data in-order,
commonly address this issue by delimiting application-level frames with
some length-value encoding,
enabling the receiver to locate the next frame in the stream
from the previous frame's position and header content.

\com{	XXX mention fixed-length option
Providing a datagram service within a TCP byte-stream
requires delimiting application records in the stream.
While TCP applications such as HTTP and SIP
traditionally delimit records
via some Type-Length-Value (TLV) encoding,
these encodings usually assume in-order delivery,
providing no reliable way for a receiving application
to locate a given record's TLV header
without having already processed
all prior TLV headers in the stream.
An exception is applications that use fixed-length records,
whose record headers always appear
at fixed multiples of the record length in the TCP stream.
To offer reliable out-of-order delivery of variable-length records, however,
\ucobs must enable the receiver to identify a record
located anywhere within a received TCP stream fragment,
{\em without} knowing the number or sizes of records
that were sent before it in the stream.
}

Since \utcp's receive path
effectively just bypasses TCP's reordering buffer,
delivering received segments to the application as they arrive,
a stream fragment received out-of-order from \utcp
may begin at any byte offset in the stream,
and not at a frame boundary meaningful to the application.
Since the receiver is by definition missing some data
sent prior to this out-of-order segment,
it cannot rely on preceding stream content
to compute the next frame's position.

\com{
If the sending application calls \verb|write()| three times
with 500 bytes each, for example,
and the sending \utcp stack transmits one segment for each \verb|write()|,
a middlebox may reassemble these segments
and then break them apart
so they arrive at the receiver as two 750-byte segments.
PEPs~\cite{rfc3234} and traffic normalizers~\cite{handley01network}
routinely re-segment TCP flows,
since they often operate by terminating the sender's original TCP stream
and forwarding received data on a separate TCP stream,
whose segment boundaries will be defined by the middlebox's
TCP stack and MTU and not the original sender's.
Such re-segmentation in the network
does not break or even affect conventional TCP applications,
so to achieve our first goal above in Section~\ref{sec:goals},
re-segmentation should not break a \utcp application either.
}

Reliable use of \utcp, therefore,
requires that frames embedded in the TCP stream be {\em self-delimiting}:
recognizable without knowledge of preceding or following data.
\abbr{	
A simple solution is to make frames fixed-length,
so the receiver can compute the start of the next frame
from the stream offset \utcp provides with out-of-order segments.
\ucobs is intended to offer a general-purpose datagram substrate, however,
and many applications require support for variable-length frames.

}{	
There are two common ways to make frames in a byte stream self-delimiting:
the sender and receiver can agree on frame boundaries in advance,
or arrange for some distinguished byte or sequence
to appear in the stream {\em only} at frame boundaries.

\paragraph{Fixed-Length Frames:}
The first approach is practical atop \utcp
if all application frames are fixed size,
since \utcp's receive path prepends a header to each received segment
containing segment's logical byte offset in the TCP stream.
If \utcp delivers a fragment with stream offset $s$,
and all frames are $f$ bytes in size,
the next frame boundary starts $f-((s-1)\ mod\ f)-1$ bytes
into the received fragment.

Most applications expect a datagram substrate
to support variable-size frames, however.
Even in media applications,
whose codecs may support a constant-bit-rate (CBR) mode,
variable- and average-bit-rate modes are commonly desired.
Even teleconferencing applications that use CBR mode
often detect and encode ``silence'' in special, small frames.
To avoid wasting bandwidth by padding all frames to a fixed size,
we wish \ucobs to support variable-size frames.

\paragraph{Variable-Length Frames:}
}
If the application-level frames happen to be encoded
so as never to include some ``reserved'' byte value, such as zero,
then we could use that byte reserved value to delimit frames
within \utcp streams.
Since we wish \ucobs to support general-purpose delivery of datagrams
of variable length containing arbitrary byte values,
however,
\ucobs must explicitly (re-)encode the application's datagrams
in order to reserve some byte value to serve as a delimiter.

Any scheme that encodes arbitrary byte streams
into strings utilizing fewer than 256 symbols will serve this purpose,
such as the ubiquitous {\em base64} scheme,
which encodes byte streams into strings
utilizing only 64 ASCII symbols plus whitespace.
Since base64 encodes three bytes into four ASCII symbols, however,
it expands encoded streams by a factor of $4/3$,
incurring a 33\% bandwidth overhead.
Since \ucobs needs to reserve only {\em one} byte value for delimiters,
and not the large set of byte values considered ``unsafe''
in E-mail or other text-based message formats,
base64 encoding is unnecessarily conservative for \ucobs' purposes.

\subsection{Operation of \ucobs}

To encode application datagrams efficiently,
\ucobs employs {\em consistent-overhead byte stuffing},
or COBS~\cite{cheshire97consistent}.
COBS is analogous to base64,
except that it encodes byte streams
to reserve only {\em one} distinguished byte value (e.g., zero),
and utilizes the remaining 255 byte values in the encoding.
COBS could in effect be termed ``base255'' encoding.
By reserving only one byte value,
COBS incurs an expansion ratio of at most 255/254,
or 0.4\% bandwidth overhead.

\com{
COBS~\cite{} is a binary encoding
ideally suited to our needs,
which eliminates {\em exactly} one byte value from a record's encoding
with minimal bandwidth overhead.
To encode an application record,
COBS first scans the record for {\em runs}
of contiguous marker-free data followed by exactly one marker byte.
COBS then removes the trailing marker,
instead {\em prepending} a non-marker byte indicating the run length.
A special run-length value indicates a run of 254 bytes
{\em not} followed by a marker in the original data,
enabling COBS to divide arbitrary-length runs into 254-byte runs
encoded into 255 bytes each,
yielding a worst-case expansion of only 0.4\%.
}

\paragraph{Transmission:}
When an application sends a datagram,
\ucobs first COBS-encodes the datagram
to remove all zero bytes.
\ucobs then prepends a zero byte to the encoded datagram,
appends a second zero byte to the end,
and writes the encoded and delimited datagram to the TCP socket.
Since this sender-side encoding and transmission process
operates entirely at application level within \ucobs,
and does not rely on any OS-level extensions on the sending host,
\ucobs operates even if the sender-side OS does not support \utcp.

The application can assign priorities
to datagrams it submits to \ucobs, however,
and if the sender's OS does support the \utcp extensions
in Section~\ref{sec:utcp-send},
\ucobs passes these priorities to the \utcp sender,
enabling higher-priority datagrams
to pass lower-priority datagrams already enqueued.
Since \utcp respects application \verb|write()| boundaries
while reordering the send queue,
\ucobs preserves its delimiting invariant
simply by writing each encoded datagram---%
with the leading and trailing zero bytes---%
in a single write.

\paragraph{Reception:}
At stream creation time,
\ucobs enables \utcp's receive-side extensions if available.
If the receive-side OS does not support \utcp,
then \ucobs simply falls back on the standard TCP API,
receiving, COBS-decoding, and delivering datagrams to the application
in the order they appear in the TCP sequence space.
(This may not be the application's original send order
if the send-side OS supports \utcp.)

If the receive-side OS supports \utcp,
then \ucobs receives segments from \utcp in whatever order they arrive,
then fits them together using the metadata in \utcp's headers
to form contiguous fragments of the TCP stream.
The arrival of a TCP segment can cause \ucobs to create a new fragment,
expand an existing fragment at the beginning or end,
or ``fill a hole'' between two fragments and merge them into one.
The portion of the TCP stream
before the receiver's cumulative-acknowledgment point,
containing no sequence holes,
\ucobs treats as one large ``fragment.''
\ucobs scans the content of any new, expanded, or merged fragment
for properly delimited records
not yet delivered to the application.
\ucobs identifies a record by the presence of two marker bytes
surrounding a contiguous sequence of bytes
containing no markers or holes.
Once \ucobs identifies a new record,
it strips the delimiting markers,
decodes the COBS-encoded content to obtain the original record data,
and delivers the record to the application.

\subsection{Why Two Markers Per Datagram?}

\begin{figure}[t]
\centering
\includegraphics[width=0.49\textwidth]{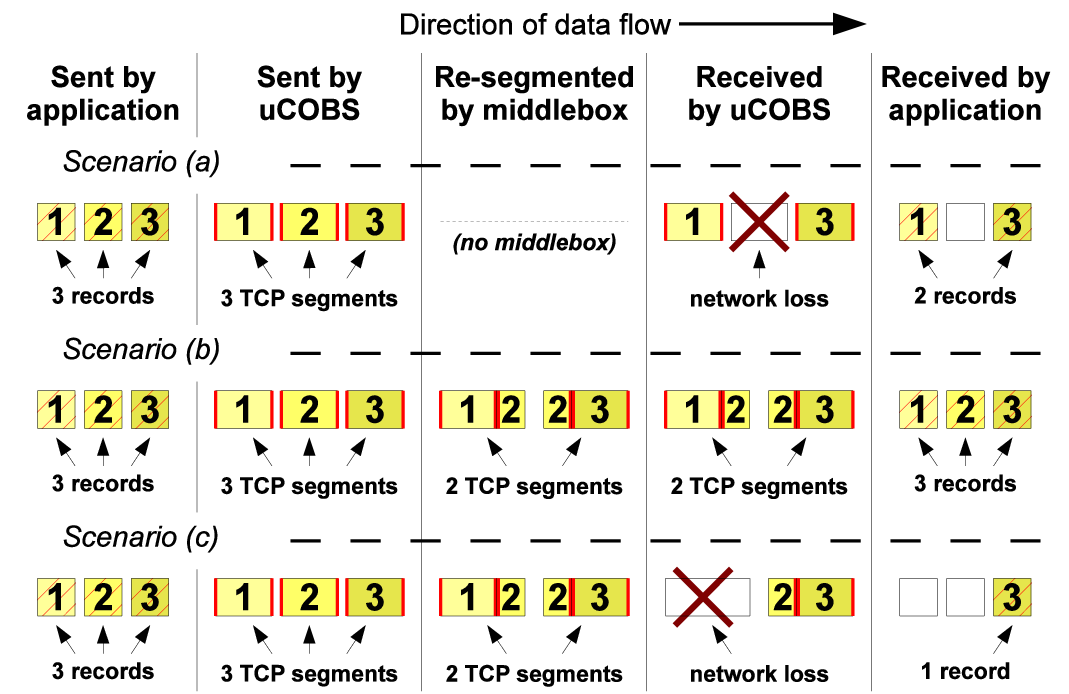}
\caption{An example illustrating a \ucobs transfer}
\label{f:tcp-minion}
\end{figure}

For correctness alone,
\ucobs need only prepend {\em or} append a marker byte to each record---%
not both---%
but such a design could reduce performance
by eliminating opportunities for out-of-order delivery.
Consider Scenario (a) in Figure~\ref{f:tcp-minion},
in which an application sends three records.
\ucobs encodes these records and sends them via three \verb|write()| calls,
which TCP in turn sends in three separate TCP segments.
In this scenario,
no middleboxes re-segment the TCP stream in the network,
but the middle segment is lost.
If the \ucobs sender were only to {\em prepend} a marker
at the start of each record,
the \ucobs receiver could not deliver record 1 immediately on receipt,
since it cannot tell if record 1 extends
into the following ``hole'' in sequence space.
Similarly,
if the sender were only to {\em append} a marker
at the end of each record,
then \ucobs could not deliver segment 3 immediately on receipt,
since record 3 might extend
backwards into the preceding hole.
By adding markers to both ends of each record,
\ucobs ensures that the receiver can deliver each record
as soon as all of its segments arrive.

These markers enable \ucobs
to offer reliable out-of-order delivery
even if network middleboxes re-segment the TCP stream.
In Scenario (b) in Figure~\ref{f:tcp-minion}, for example,
\ucobs sends three records encoded into three TCP segments as above,
but a middlebox re-segments them into two longer TCP segments,
whose boundary splits record 2 into two parts.
If neither of these segments are lost,
then the \ucobs receiver can deliver record 1
immediately upon receipt of the first TCP segment,
and can deliver records 2 and 3
upon receipt of the second segment.
If the first segment is lost
as shown in Scenario (c), however,
the \ucobs receiver cannot deliver the missing record 1
or the partial record 2,
but can still deliver record 3 as soon as the second TCP segment arrives.



\section{\utls: Secure Datagrams on TCP}
\label{sec:utls}

\com{
\begin{itemize}
  \item TLV breaks because of synchronization issue;  
    but this can be fixed with hashes.
  \item describe, with figures, SSL minion design.
\end{itemize}
}

While \ucobs offers out-of-order delivery
wire-compatible up to the TCP level,
middleboxes often inspect and manipulate
the {\em content} of TCP streams as well~\abcite{
	reis08detecting}{vratonjic10integrity}.
All unencrypted network traffic today is, {\em de facto},
``fair game'' for middleboxes---%
and streams exhibiting
any ``out of the ordinary'' middlebox-visible behavior
are likely to fail over {\em some} middleboxes~\cite{
	honda11possible}.
An application's only way to protect ``end-to-end'' communication in practice,
therefore,
is via end-to-end encryption and authentication.
But network-layer mechanisms such as IPsec~\cite{rfc4301}
face the same deployment challenges as new secure transports~\cite{
	ford07structured},
and remain confined to the niche of corporate VPNs.
Even VPNs are shifting from IPsec toward HTTPS tunnels~\cite{
	davies09directaccess},
the only form of end-to-end encrypted connection
almost universally supported on today's Internet.
A network administrator or ISP might disable nearly any other port
while claiming to offer ``Internet access,''
but would be hard-pressed to disable HTTPS,
today's foundation for E-commerce.

We could layer TLS atop \ucobs,
but TLS decrypts and delivers data only in-order,
negating \utcp's benefit.
We could also layer the datagram-oriented DTLS~\cite{rfc4347}
atop \ucobs,
but the resulting
(DTLS-encrypted {\em then} COBS-encoded) wire format
would be radically different from TLS over TCP\@,
and likely fail to traverse middleboxes expecting TLS,
particularly on port 443.

The goal of \utls, therefore,
is to coax out-of-order delivery from
the {\em existing} TCP-oriented TLS wire format,
producing an encrypted datagram substrate
indistinguishable on the wire from standard TLS connections
(except via analysis of ``side-channels''
such as packet length and timing,
which we do not address).
Run on port 443,
a \utls stream is indistinguishable from HTTPS---%
regardless of whether the application actually uses HTTP headers,
since the HTTP portion of HTTPS streams are TLS-encrypted anyway.
Deployed this way,
\utls effectively offers an end-to-end protected substrate
in the ``HTTP as the new narrow waist'' philosophy~\cite{popa10http}.

\subsection{Design of \utls}

TLS~\cite{rfc5246} already breaks its communication into {\em records},
encrypts and authenticates each record,
and prepends a header
for transmission on the underlying TCP stream.
TLS was designed to decrypt records strictly in-order, however,
creating three challenges for \utls:

\begin{itemize}
\item	{\bf Locating record headers out-of-order.}
	Since encrypted data may contain any byte sequence,
	there is no reliable way to differentiate a TLS header
	from record data in the TCP stream,
	as COBS encoding provides.
\item	{\bf Encryption state chaining.}
	Some TLS ciphersuites
	chain encryption state across records,
	making records indecipherable
	until prior records are processed.
\item	{\bf Record numbers used in MAC computation.}
	TLS includes a record number,
	which increases by 1 for each record,
	in computing the record's MAC.
	But the \utls receiver may not know an out-of-order record's number:
	holes in TCP sequence space before the record
	could contain an unknown number of prior records.
\end{itemize}

To locate records out-of-order,
\utls first scans a received stream fragment for byte sequences
that {\em may} represent the TLS 5-byte header:
i.e., containing the correct record type and version,
and a plausible length.
While this scan may yield false positives,
\utls verifies the inferred header
by attempting to decrypt and authenticate the record.
If the cryptographic MAC check fails,
instead of aborting the connection as TLS normally would,
\utls assumes a false positive and continues scanning.

Since TLS's MAC is designed to prevent resourceful adversaries
from constructing a byte sequence
the receiver could misinterpret as a record,
and it is by definition at least as hard
to find such a sequence ``accidentally'' as to forge one maliciously,
TLS security should protect equally well
against accidental false positives.
One exception is when TLS is using its ``null ciphersuite,''
which performs no packet authentication.
With this ciphersuite,
normally used only during initial key negotiation,
\utls disables out-of-order delivery to avoid the risk
of accepting and delivering false records.

The only obvious solution to the second challenge above
is to avoid ciphersuites
that chain encryption state across records.
Most ciphersuites before TLS 1.1 chain encryption state, unfortunately.
Any stream cipher inherently does so,
such as the RC4 cipher used in early SSL versions.
Most recent ciphersuites use block ciphers in CBC mode.
CBC ciphers do not inherently depend on chained encryption state,
but do require an Initialization Vector (IV) for each record.
Until recently, TLS produced each record's IV implicitly
from the prior record's encryption state,
making records interdependent.

To fix a security issue, however,
TLS 1.1 block ciphers use explicit IVs,
which the sender generates independently for each record
and prepends to the record's ciphertext.
As a side-effect, TLS 1.1 block ciphers support out-of-order decryption.
Since TLS supports negotiation of versions and ciphersuites,
\utls simply leverages this process.
An application can insist on TLS 1.1 with a block cipher
to ensure out-of-order delivery support,
or it can permit older ciphersuites
to maximize interoperability,
at the risk of sacrificing out-of-order delivery.

The third challenge is the implicit ``pseudo-header''
TLS uses in computing the MAC for each packet.
This pseudo-header includes a ``sequence number''
that TLS increments once per {\em record},
rather than per {\em byte} as with TCP sequence numbers.
When \utls identifies a possible TLS record
in a TCP fragment received out-of-order,
the receiver knows only
the byte-oriented TCP stream offset,
and not the TLS record number.
Since records are variable-length,
unreceived holes prior to a record to be authenticated
may ``hide'' a few large records or many smaller records,
leaving the receiver uncertain
of the correct record number for the MAC check.

To authenticate records out-of-order
without modifying the TLS ciphersuite,
therefore,
\utls attempts to {\em predict} the record's likely TLS record number,
using heuristics such as the average size of past records,
and may try several adjacent record numbers
to find one for which the MAC check succeeds.
If \utls fails to find a correct TLS record number,
it cannot deliver the record out-of-order,
but will still eventually deliver the record in-order.

\label{sec:utls-send}
The current \utls supports only receiver-side unordered delivery,
and not the send-side \utcp enhancements in Section~\ref{sec:utcp-send},
because send-side reordering complicates record number prediction.
A future enhancement we intend to explore
is for \utls to prepend an explicit record number to application payloads
before encryption.
Since encryption does not depend on record number,
the receiving \utls can decrypt the record number for use in the MAC check,
avoiding the need to predict record numbers
and enabling send-side reordering.
Since the only wire-protocol change is protected by encryption,
the change would be invisible to middleboxes.
Preserving end-to-end backward compatibility
may require a way to negotiate ``under encryption''
the use of explicit record numbers, however.

\com{
First, the application messages must be encrypted at the sender.  SSL Minion
does the encryption {\em after} the COBS encoding.  For example, TCP
Minion performs COBS on the application message, removing zeroes and
bookending with a zero byte.  This encoded message is then encrypted using SSL,
producing a {\em record}, which is passed to the underlying Kernel TCP
Stream.

The second additional component required by SSL Minion is the decryption at the
receiver.  In TCP
Minion, the receiver passes the application a message whenever it sees two
zeroes and a ``hole-less'' data stream between them.  For SSL Minion, that
check cannot be performed until a complete SSL record has been received and
decrypted.  Thus, the receiver breaks the Kernel TCP Stream into SSL records. 
SSL Minion decrypts complete (i.e., hole-less) SSL records, before passing them
to TCP Minion for COBS decoding.  COBS decoding may or may not occur
immediately, depending on whether the stream contains an application message. 
Finally, the decoded message, now containing its original application data, is
passed to the application.

Just as in regular TCP Minion, SSL Minion passes messages to
the application in an {\em order-agnostic} way.  Because of this, stream ciphers
cannot be used because they rely on previous ssl records for encryption and
decryption.  This prevents decryption of a record until after receiving previous
records.  Furthermore, not even all block ciphers can be used because some block
ciphers rely on previous blocks to encrypt a given block.
}


\section{Prototype Implementation}
\label{sec:impl}


This section describes the current Minion prototype,
which implements \utcp in the Linux kernel,
and implements \ucobs and \utls in application-linked libraries.
The \utcp prototype is Linux-specific,
but we expect the API extensions it implements
and the application-level libraries
to be portable.

\com{	stated several times earlier
Note that the send- and receive-side changes for \utcp
are independent
and both sides are not required for functionality,
only for performance.
The \utcp prototype's design aims to minimize kernel changes
over other goals such as elegance, optimization, or application convenience.
}

\paragraph{The \utcp Receiver in Linux:}
The \utcp prototype
adds about 240 lines and modifies about 50 lines of code 
in the Linux 2.6.34 kernel,
to support the new \verb|SO_UNORDERED| socket option.
This extension involved two main changes.
First, 
\utcp modifies the TCP code
that delivers segments to the application,
to prepend a 5-byte metadata header to the data
returned from each \verb|read()| system call.
This header consists of a 1-byte flags field
and a 4-byte TCP sequence number.
One flag bit is currently used,
with which \utcp indicates whether it is delivering data
in-order or out-of-order.
Second,
if TCP's in-order queue is empty,
\utcp's \verb|read()| path
checks and returns data from the out-of-order queue.
To minimize kernel changes,
segments remain in the out-of-order queue after delivery,
so \utcp will eventually deliver the same data again in-order.

\paragraph{The \utcp Sender in Linux:}
On the send path,
\utcp adds about $250$ lines of kernel code and modifies about $20$
in Linux 2.6.34,
supporting a new
\verb|SO_UNORDEREDSEND| socket option
via two changes.

First,
\utcp expects the application to prepend a 5-byte header,
containing a 1-byte flags field and a 4-byte tag,
to the data passed to each \verb|write()|.
The flags are currently unused,
and
the tag indicates message priority.

Second,
\utcp inserts the data from each \verb|write()|
into the kernel's send queue in priority order.
Linux's TCP send queue 
is a simple FIFO that packs application data
into kernel buffers
sized to the TCP connection's Maximum Segment Size (MSS).
When inserting application messages non-sequentially,
however,
\utcp must preserve application message boundaries in the kernel.
For simplicity,
\utcp allocates kernel buffers (\verb|skbuff|s) so that
each message sent via \utcp starts a new \verb|skbuff|,
and may span several \verb|skbuff|s,
but no \verb|skbuff| contains data from multiple application writes.

Disabling Linux's usual packing of
MSS-sized \verb|skbuff|s can affect
Linux's congestion control, unfortunately,
which counts \verb|skbuff|s sent instead of bytes.
Section~\ref{sec:bw-cpu} discusses the effects of
this Linux-specific issue,
which a future version of \utcp will address.

\com{
To address this issue, before doing prioritized insertions, 
we first check whether the immediately prior skbuff in the queue has the same
priority as the current insertion, and
if so, we append the new write data to that skbuff (up to MSS size) instead of
always creating a brand-new skbuff. 
This doesn't hurt the application anyway,
since the application has no way to request the insertion of a new write in
between two previously-written segments of the same priority. For common case,
i.e. mostly same (low-)priority data on a stream,
the results are comparable to plain TCP as shown in Section~\ref{sec:bw-cpu}. 
}
\paragraph{The \ucobs Library:}
The \ucobs prototype is a user-space library in C,
amounting to \texttildelow$700$ lines of code~\cite{cloc-153}.
\ucobs presents simple
\verb|cobs_sendmsg()| and \verb|cobs_recvmsg()| interfaces
enabling applications to send and receive COBS-encoded datagrams,
taking advantage of send-side prioritization and out-of-order reception
depending on the presence of send- and receive-side OS support
for \utcp, respectively.

\paragraph{A \utls Prototype Based on OpenSSL:}
The \utls prototype builds on OpenSSL 1.0.0~\cite{openssl},
adding \texttildelow$550$ lines of code
and modifying \texttildelow$40$ lines~\cite{cloc-153}.
Applications use OpenSSL's normal API
to create a TLS connection atop a TCP socket,
then set a new \utls-specific socket option
to enable out-of-order, record-oriented delivery on the socket.
OpenSSL 1.0.0 unfortunately does not yet support TLS 1.1,
the first TLS version
that uses explicit Initialization Vectors (IVs),
permitting out-of-order decryption.
For experimentation, therefore,
the \utls prototype modifies OpenSSL's TLS 1.0 ciphersuite
to use explicit IVs as in TLS 1.1.
Since this change breaks OpenSSL's interoperability,
our prototype is not suitable for deployment.
We are currently
porting \utls to the next major OpenSSL release,
which supports TLS 1.1.

\com{
OpenSSL: cloc output
-------------------------------------------------------------------------------
Language                     files          blank        comment           code
-------------------------------------------------------------------------------
SUM:
 same                         1404              0          86205         322690
 modified                       12              0              0             40
 added                           0            119            181            800
 removed                         1             13             15             86
-------------------------------------------------------------------------------

Linux (w receiver side changes only):
-------------------------------------------------------------------------------
SUM:
 same                         28587              0        1885539        8432162
 modified                         9              0              6             53
 added                            0             43            223            242
 removed                          0              0              1              8
-------------------------------------------------------------------------------

SUM:                          28596        1636350        1885890        8432714

}

\section{Performance Evaluation}
\label{sec:perf}

This section evaluates Minion
via experiments designed to approximate realistic application scenarios.
All experiments were run across three Intel PCs running Linux 2.6.34:
between two machines representing end hosts,
a third machine interposes on the path
and uses dummynet~\cite{carbone10dummynet}
to emulate various network conditions.
To minimize well-known TCP delays
fairly for both TCP and \utcp,
we 
enabled Linux's ``low latency'' TCP code path
via the \verb|net.ipv4.tcp_low_latency| sysctl,
and
disabled the Nagle algorithm.

\subsection{Bandwidth and CPU Costs}
\label{sec:bw-cpu}

We first explore \utcp's costs,
with and without record encoding and extraction via \ucobs and \utls,
for a $30$MB bulk transfer on a path with $60$ms RTT.

\paragraph{Raw \utcp:}
\com{
\begin{figure}[tbp]
\centering
\includegraphics[width=0.49\textwidth]{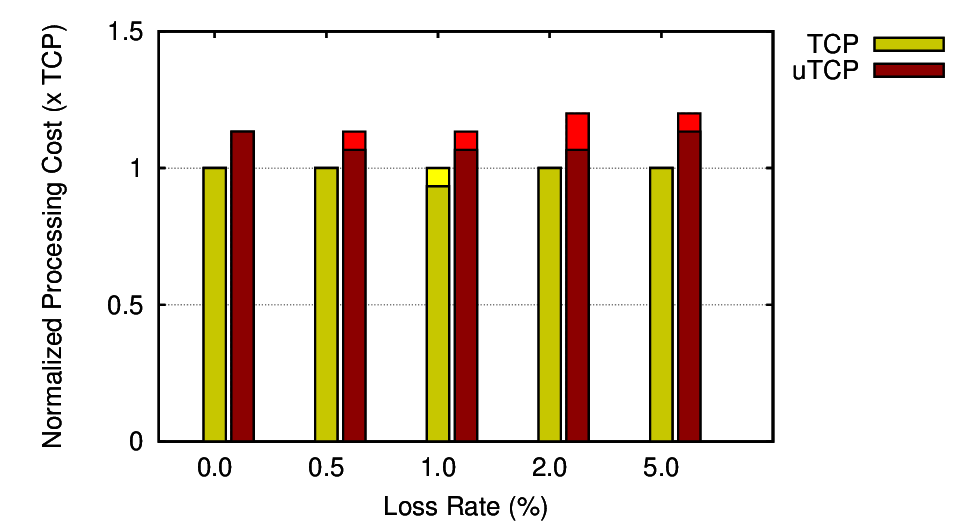}
\caption{CPU costs of application with and without ofo-send modifications at different loss rates.}
\label{f:raw-costs}
\end{figure}
}

\abbr{}{
\begin{figure}[tbp]
\centering
\includegraphics[width=0.49\textwidth]{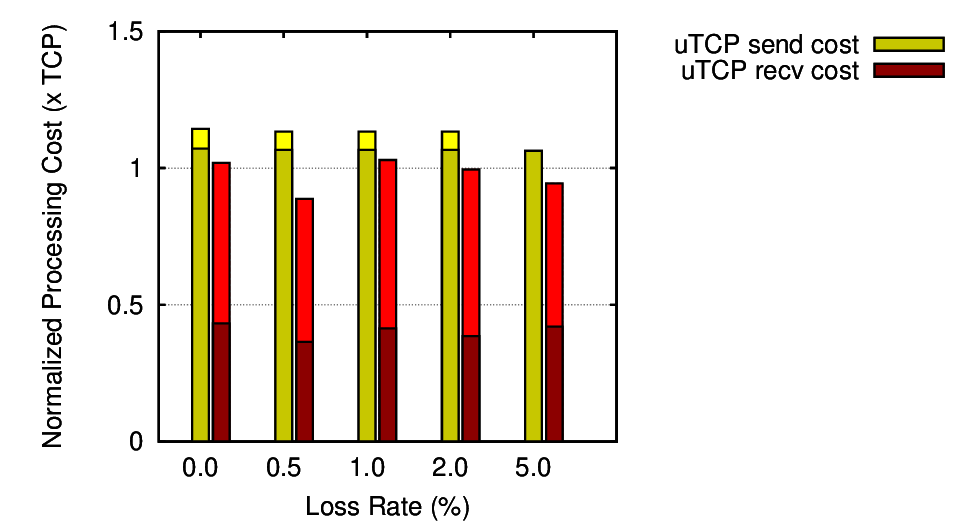}
\caption{CPU costs of application with and without ofo-send modifications at different loss rates.}
\label{f:raw-costs}
\end{figure}
}

\utcp's CPU costs
at both the sender and the receiver,
without application-level processing,
are almost identical to
TCP's CPU costs,
across a range of loss rates
from 0\% to 5\%
\abbr{
(figure omitted for space reasons).
}{
(Figure~\ref{f:raw-costs}).
}

\begin{figure}[tbp]
\centering
\includegraphics[width=0.49\textwidth]{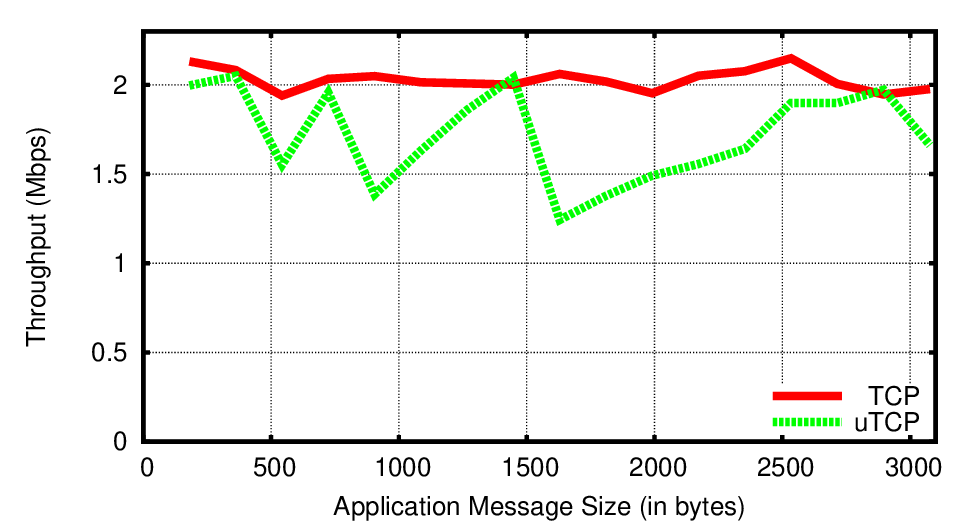}
\caption{Throughput with different app message sizes.}
\label{f:msgsize-costs}
\end{figure}
\com{
{\bf GRAPH TODO: Average over 3 runs at least, and run with utcp on both sides. 
  ON THE SAME GRAPH, plot three curves: 
  normalized throughput, 
  normalized processing cost at receiver,
  and normalized processing cost at sender.}
}
Figure~\ref{f:msgsize-costs}
shows bandwidth achieved
by raw \utcp and TCP,
for different application \verb|write()| sizes.
When the message size is a multiple of TCP's Maximum Segment Size (MSS)---%
at $1448$ bytes ($1$ MSS) and $2896$ bytes ($2$ x MSS)---%
\utcp's throughput is the same as TCP's.

The disparity elsewhere is due to Linux's congestion control
counting \verb|skbuff|s instead of bytes,
mentioned earlier in Section~\ref{sec:impl}. 
We partially address this issue
by coalescing data into \verb|skbuff|s where easily possible.
More specifically, we coalesce small messages
when they fully fit within MSS-sized \verb|skbuff|s
at the tail of the sender-side socket buffer.
This fix makes \utcp throughput similar to TCP's when
the MSS is divisible by message size---%
at $362$ bytes ($\frac{1}{4}$ MSS) and $724$ bytes ($\frac{1}{2}$ MSS).
Future versions of \utcp
will fully address this Linux-specific issue
with changes either to \utcp or to Linux's congestion control.

\com{
For messages just larger than 1/2 MTU, the kernel 
can't pack multiple app messages in 
one MTU without preventing out-of-order delivery and thus for 
a given number of \verb|skbuff|s
\utcp transmits fewer bytes than TCP,
reducing throughput.

, before doing prioritized insertions, 
we first check whether the immediately prior skbuff in the queue has the same
priority as the current insertion, and
if so, we append the new write data to that skbuff (up to MSS size) instead of
always creating a brand-new skbuff. 
This doesn't hurt the application anyway,
since the application has no way to request the insertion of a new write in
between two previously-written segments of the same priority. For common case,
i.e. mostly same (low-)priority data on a stream,
the results are comparable to plain TCP as shown in Section~\ref{sec:bw-cpu}. 
}
\abbr{
}{

\com{
The throughput disparity between TCP and \utcp at the other message sizes is 
a result of negative interference between two implementation choices:
the Linux TCP implementation uses segment-based congestion control,
and assumes that every kernel memory block (\verb|skbuff|) in the sender's queue is MSS-sized.
As discussed in Section~\ref{sec:impl},
\utcp limits the size of the \verb|skbuff|s on the send-queue
when the application message size is not aligned on MSS-sized boundaries
to preserve message boundaries
and to enable non-sequential insertion of application data in the queue.
Thus, for the same number of \verb|skbuff|s, 
each of which results in a separate TCP segment,
\utcp transmits fewer bytes than TCP does
when the app message size is not a multiple of MSS,
resulting in reduced throughput.
}

While we believe that the correct way to fix this negative interference
is to change the Linux congestion control mechanisms to be byte-based,
as is done in BSD for instance,
this is a particular implementation artifact
that we do not address further in this paper.
In our experiments,
we therefore use MSS-sized ($1448$-byte) application messages
where possible;
where it is unavoidable,
our results include this throughput penalty
when the \utcp sender is used,
and should improve when this interference is eliminated.
}

\paragraph{Costs with \ucobs/\utls:}
To measure these CPU costs, 
we run
a 30MB bulk transfer
over a path with a 60ms RTT, 
for several loss rates.

\begin{figure}[tbp]
\centering
\includegraphics[width=0.49\textwidth]{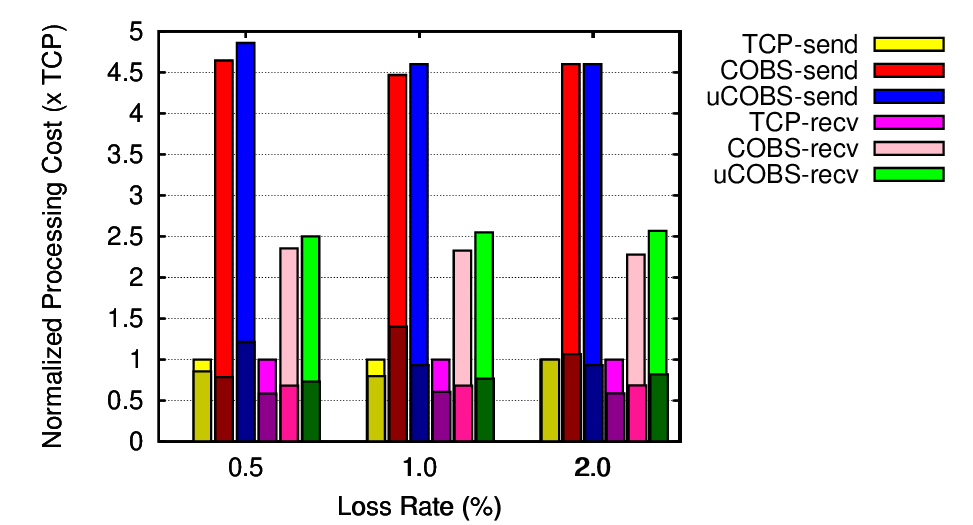}\\
(a) COBS/uCOBS encoding costs\\
\includegraphics[width=0.49\textwidth]{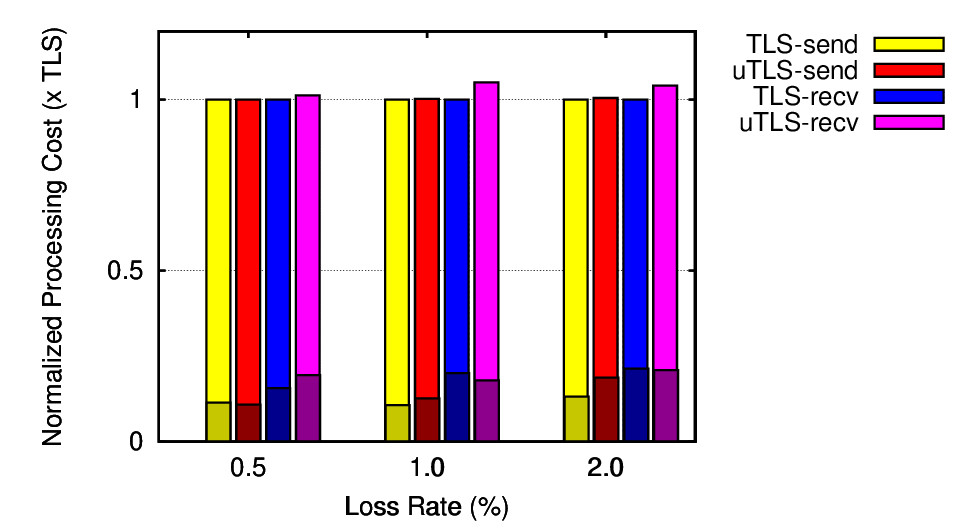}\\
(b) TLS/uTLS costs
\caption{CPU costs of using an application with TCP, COBS, and uCOBS
at different loss rates. }
\label{f:costs}
\end{figure}

Figure~\ref{f:costs}(a) shows CPU costs
including application-level encoding/decoding,
atop standard TCP (``COBS'')
and atop \utcp (``\ucobs''),
for several loss rates
at both sender and receiver.
The lighter part of each bar represents user time
and the darker part represents kernel time.
These results are normalized to the performance
of raw TCP,
with no application-level encoding or decoding.

\com{
these user-space libraries represent
only two of the many ways applications might utilize \utcp,
and that these libraries were written
with little emphasis on tuning or optimization.
}

COBS encoding/decoding
barely affects kernel CPU use
but incurs some application-level CPU cost.
This cost
is partly due to the encoding itself,
and partly because the libraries are not yet well-optimized.

Figure~\ref{f:costs}(b) shows the CPU costs
of \utls relative to TLS.
At the sender,
the CPU costs
are identical, 
since there is nothing that \utls does differently 
than TLS, 
and since the CPU cost of using \utcp 
is practically the same as with TCP.
The user-space cost for
the \utls receiver
is generally higher than TLS,
since the \utls receiver does more work in processing out-of-order frames
than the TLS receiver,
but this cost remains within 7\% of the TLS receiver's cost.



The bandwidth penalty of \ucobs encoding
is barely perceptible, under 1\%.
TLS's bandwidth overhead, up to 10\%,
is due to TLS headers, IVs, and MACs;
\utls adds no bandwidth overhead beyond standard TLS 1.1.

\com{
Figure~\ref{f:costs}(a) shows
a stacked bar graph:
the total height of the stacked bar represents
total CPU time consumed,
the darker component is CPU time in kernel mode,
and the lighter component is CPU time in user mode.
COBS and \ucobs both consume more system time
since the COBS library
uses \verb|malloc()| and \verb|free()| system calls
to maintain its reassembly queue in user space.

Bandwidth costs are minimal;
as expected, 
TLS and \utls have the highest bandwidth overhead 
due to the overhead added by the TLS header.
}

\subsection{Conferencing Applications}

\begin{figure}[tbp]
\centering
\includegraphics[width=0.49\textwidth]{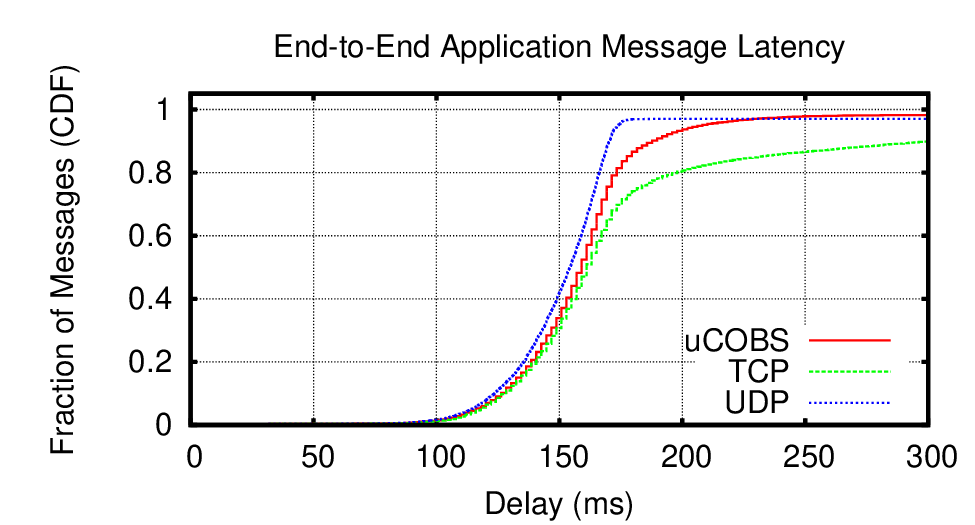}
\caption{CDF of end-to-end latency in VoIP frames.}
\label{f:voip-latency}
\end{figure}

\abbr{ 
\begin{figure}[tbp]
\centering
\includegraphics[width=0.49\textwidth]{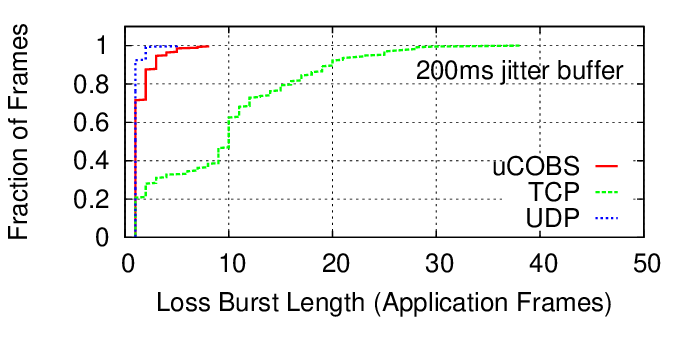}
\caption{CDF of codec-perceived loss-burst size with TLV encoded frames over TCP, UDP, and uCOBS.}
\label{f:voip-burst-loss}
\end{figure}

\begin{figure*}[tbp]
\centering
\includegraphics[width=0.99\textwidth]{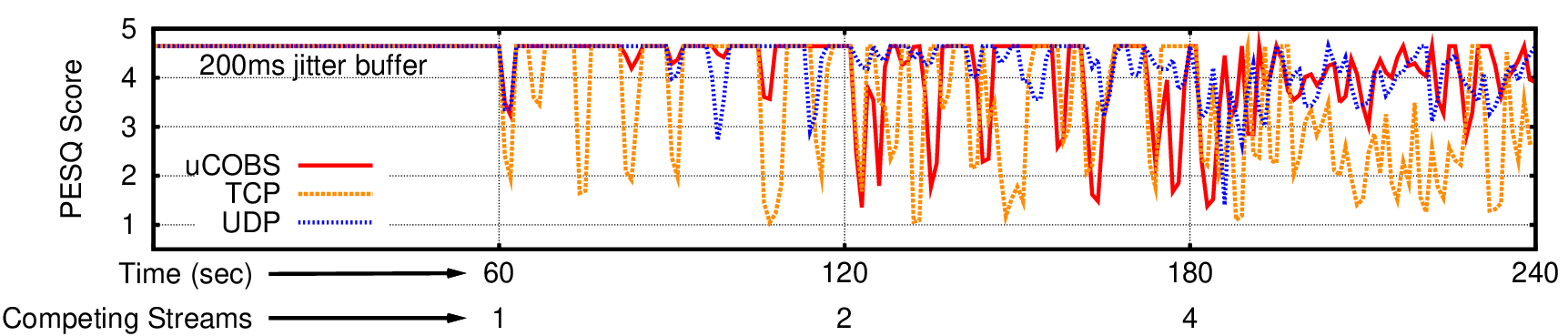}
\caption{Moving PESQ score of VoIP call under increasing bandwidth competition.}
\label{f:voip-pesqstk}
\end{figure*}
}{ 
\begin{figure}[tbp]
\centering
\includegraphics[width=0.49\textwidth]{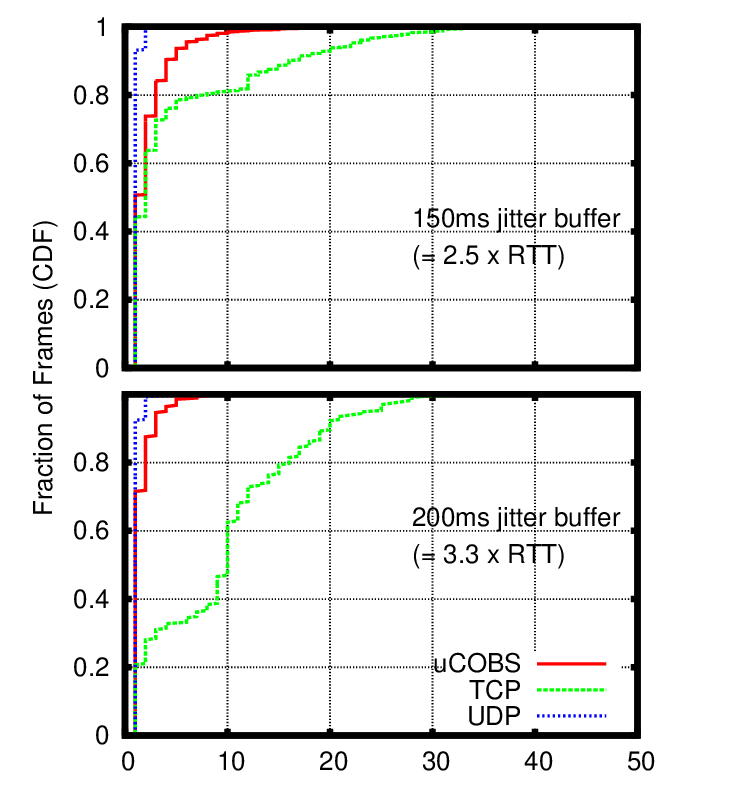}
\caption{CDF of codec-perceived loss-burst size with TLV encoded frames over TCP, UDP, and uCOBS.}
\label{f:voip-burst-loss}
\end{figure}

\begin{figure*}[tbp]
\centering
\includegraphics[width=0.99\textwidth]{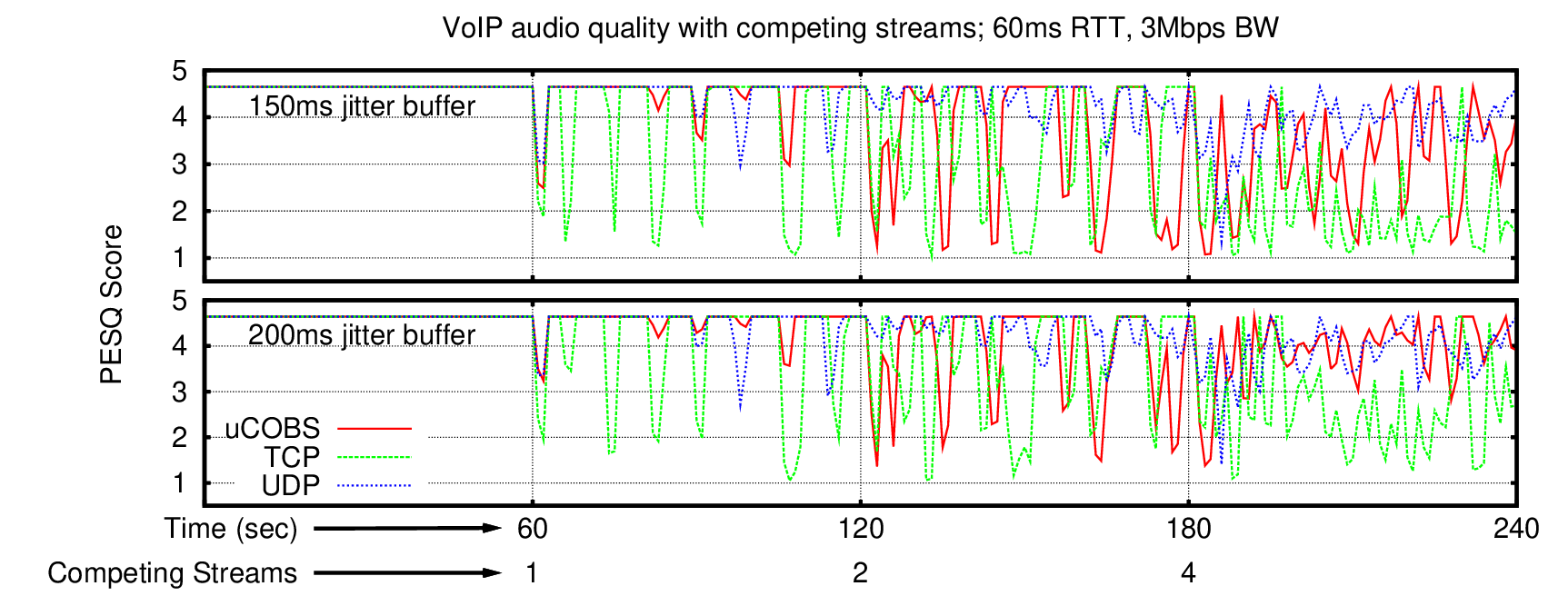}
\caption{Moving PESQ score of VoIP call under increasing bandwidth competition.}
\label{f:voip-pesqstk}
\end{figure*}
} 

We now examine a real-time Voice-over-IP (VoIP) scenario.
A test application uses the SPEEX codec~\cite{speex}
to encode a WAV file using ultra-wideband mode ($32$kHz),
for a $256$kbps average bit-rate,
and transmit voice frames at fixed $20$ms intervals.
Network bandwidth is $3$Mbps and RTT is $60$ms,
realistic for a home broadband connection.
To generate losses more realistically representing network contention,
we run a varying number of competing TCP file transfers,
emulating concurrent web browsing sessions or a BitTorrent download,
for example.

This is a simplistic scenario for experimental purposes.
Real VoIP applications,
which we intend to evaluate in future work,
often determine bit-rate based on network conditions.
Real applications may also implement loss recovery mechanisms atop UDP,
which may improve perceived voice quality when using UDP.

\paragraph{Latency:}

Figure~\ref{f:voip-latency}
shows a CDF of one-way per-frame latency
perceived by the receiving application,
\com{under heavy contention from $10$ competing TCP streams.
With \ucobs over \utcp,
about $40$\% of frames are significantly delayed due to retransmissions.
$10$\% of transmitted UDP frames
do not arrive at all,
since UDP does not retransmit.
With standard TCP,
an additional $20$\% of frames experience significant delays
due to in-order delivery.
With less network contention,
latency differences are less pronounced,
but the curves remain similar.}
under heavy contention from $4$ competing TCP streams.
All three transports suffer major delays.
$4$\% of UDP frames 
do not arrive at all,
since UDP does not retransmit.
$95$\% of frames sent with \ucobs over \utcp
arrive within $200$ms,
compared to $80$\% of TCP frames.

\paragraph{Burst Losses:}

VoIP codecs such as SPEEX
can interpolate across one or two missing frames,
but are sensitive to burst losses or delays,
which yield user-perceptible blackouts.
An application's susceptibility to blackouts
depends on its jitter buffer size:
a larger buffer increases the receiver's tolerance of burst losses or delays,
but also increases effective round-trip delay,
which can add user-perceptible ``lag'' to all interactions.

The CDF in Figure~\ref{f:voip-burst-loss}
shows the prevalence of different lengths of burst losses
experienced by the receiver in a typical VoIP call.
A burst loss is a series of consecutive voice frames
that miss their designated playout time,
due either to loss or delay.
\abbr{
}{
A $150$ms jitter buffer and $4$ competing streams
atop TCP appears similar to \ucobs and UDP,
but this is misleading.
TCP has $3\times$ as many {\em total bursts} with a $150$ms
jitter buffer as with $200$ms,
with most of the extra bursts only a single packet.
If these single dropped packets are kept by
a larger jitter buffer,
as in the bottom graph,
TCP shows a much bleaker picture.
}

A $200$ms jitter buffer of $3\times$ the path RTT might seem generous,
but the ITU's recommended maximum transmission 
time of $400$ms~\cite{itu03transtime}
allows for a larger buffer with these network conditions.
Now the differences between \ucobs and TCP are quite pronounced,
with $80$\% of burst losses atop \ucobs being $3$ or fewer packets,
nearly matching that of UDP.
Meanwhile $40$\% of TCP's bursts are greater than $10$ packets,
producing highly-perceptible $1/5$-second pauses.

\paragraph{Perceptual Audio Quality:}

\abbr{
To illustrate the impact of unordered delivery on VoIP quality,
we use Perceptual Evaluation of Speech Quality (PESQ)~\cite{itu07wideband}
to measure audio reproduction quality,
by comparing the audio stream reproduced by SPEEX at the receiver
against that of an ideal run with no lost or delayed frames.
We transmit a $4$-minute VoIP call
using a jitter buffer of $200$ms,
introducing 1 to 4 competing TCP streams progressively
at $1$-minute intervals.

Figure~\ref{f:voip-pesqstk}
plots PESQ quality scores for 2-second sliding time windows
over a representative 4-minute call,
comparing transmission via \ucobs, TCP, and UDP.
The effect of network contention becomes apparent
even with only one competing stream,
but unordered delivery makes this impact
much smaller on \ucobs or UDP than on TCP.
\ucobs sometimes performs better than UDP, in fact,
when \utcp successfully retransmits a lost segment
within the jitter buffer's time window,
whereas UDP never retransmits.
(Some UDP applications employ application-level retransmission
schemes~\cite{openarena}, especially for control data.)
Like TCP, \ucobs shows greater volatility than UDP with higher contention,
due to TCP congestion control effects that \utcp preserves
(though congestion control can be disabled).
Similarly, the ``back-off'' of the {\em competing} streams
enables the transports to rebound after the initial
contention of 4 competing streams.

}{	
The next experiment
shows objectively that the characteristics demonstrated above for TCP,
\utcp and UDP directly affect the perceived audio quality of a VoIP call.
The Perceptual Evaluation of Speech Quality (PESQ)~\cite{itu07wideband} is an 
industry-recognized algorithm for objectively measuring the quality 
of a degraded sound file compared to a target sound file.
In this work,
we use the freely available International Telecommunication Union (ITU)
reference implementation~\cite{itu05impl} for 862.2 PESQ score.
PESQ scores range from roughly $1$ (degraded audio is inaudible) to $4.644$
(perfect match, no degradation).

We transmit a $4$ minute VoIP call using the 
same $60$ms RTT and $3$Mbps link as before,
with two jitter buffer sizes.
Competing streams begin at $1$-minute intervals,
eventually causing significant degradation of 
audio quality with $4$ competing streams.

We compare the received audio files to 
an ``ideal'' audio file with no dropped frames.
Each data point represents the PESQ score for the next $2$ seconds.
\xxx{ Why do we choose $2$ seconds?}
We arbitrarily choose $2$ seconds because
we feel this is a reasonable amount of time a user considers
when answering the question: ``How is audio quality now?''.

Our testing shows that audio quality degrades
immediately once competing streams begin.
Our goal is to show the user experience throughout 
the call as competing streams choke out the bandwidth.
In such a scenario,
we expect to see the audio quality rebound after the
initial collision with competing TCP streams,
which themselves back off due to their losses.
Thus,
Figure~\ref{f:voip-pesqstk} simulates a more
sustained user experience over the course of a $4$ minute call,
and represents a typical result for this experiment.

\com{
\begin{figure}[tbp]
\centering
\includegraphics[width=0.49\textwidth]{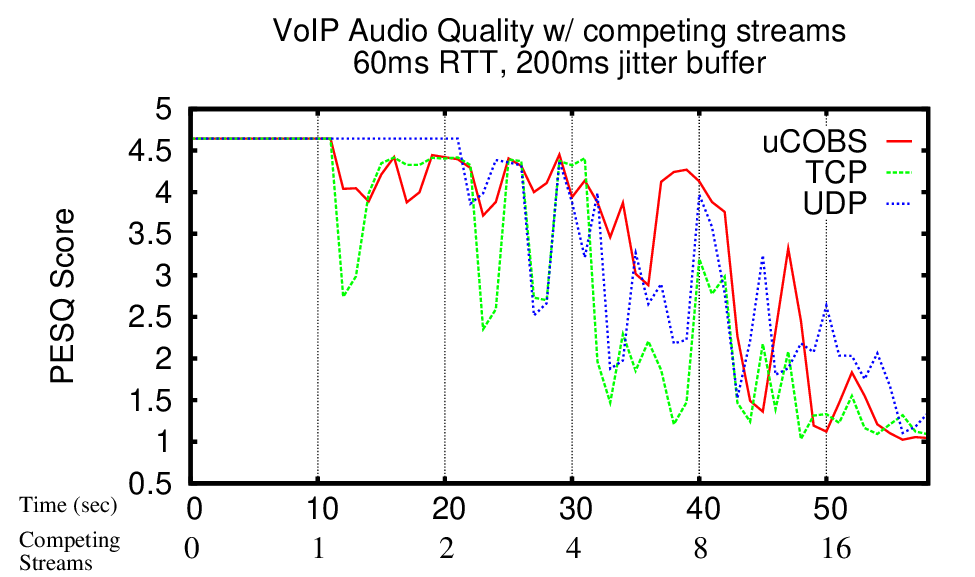}
\caption{Moving PESQ Score of VoIP audio quality under increasing bandwidth 
competition.}
\label{f:voip-pesq}
\end{figure}

Figure~\ref{f:voip-pesq} shows the moving PESQ score,
measured every second,
over the duration of a $1$ minute VoIP call.
The figure shows that \ucobs consistently delivers better audio 
quality than TCP,
while approximating or bettering that of UDP at most points in the call.
}

We see that even $1$ competing stream causes significant drops for TCP,
while \ucobs and UDP exhibit only minor dips in quality.
In the third minute of the call ($2$ competing streams),
\ucobs shows volatility compared to UDP\@.
We attribute this to the congestion control of \ucobs and TCP\@.
The competing streams themselves use TCP's backoff mechanism,
interleaving windows of high and low bandwidth for the VoIP call.
\ucobs still suffers from this less than TCP,
however,
due to its ability to deliver data out-of-order.
Meanwhile,
TCP's decreasing fidelity beyond $1$ competing stream makes it 
potentially a non-option under any bandwidth competition.

Perhaps the most important feature shown by these experiments is the effect of
the jitter size on \ucobs,
and the lack of effect for TCP.
Increasing the jitter buffer from $150$ to $200$ms,
particularly under $4$ competing streams,
improves \ucobs significantly more than it improves TCP.
This speaks to the fundamental difference between the two
highlighted in Figure~\ref{f:voip-burst-loss}:
that TCP's bursty losses are ill-suited to real-time applications.
Because a single lost packet delays subsequent packets,
repeatedly dropped packets under heavy bandwidth contention
are helpless with any reasonable jitter buffer size.
Breaking the in-order requirement of delivery is essential to transmission
quality in these situations,
making \ucobs more suitable for real-time applications.
Indeed,
increasing the jitter buffer size to $200$ms brings \ucobs 
in-line with UDP under $4$ competing streams.

\com{
Take inspiration from "The Delay-Friendliness of TCP" paper

Different scenarios:
- CBR or ABR Voice codec versus Voice+Video codecs (e.g., SPEEX)
- a couple different sampling rates
- Different network conditions: loss rates?  number of competing streams?
- Jitter buffer size - large for streaming/VoD, small for conferencing
- TCP minion versus SSL minion

In each graph, compare TCP versus TCPOO versus UDP?

Graph {\bf (Done. See Figure~\ref{f:voip-latency})}:
key point: use of TCPOO reduces delay for many packets
one-way latency/jitter CDF.
a "more realistic" version of the latency graph in 7.1.
PDF or CDF:
X axis: one-way latency (or latency difference above minimal)
Y axis: percentage of VoIP frames

Graph {\bf (Done.  See Figure~\ref{f:voip-burst-loss}.  
60ms RTT, 10Mbps BW, 10 competing TCP flows, and 40ms jitter buffer.)}:
key point: use of TCPOO greatly reduces the incidences and lengths
of bursty losses from perspective of codec
PDF:
X axis: codec-perceived too-late burst length (number of frames)
Y axis: count or percentage
}

\com{
Graph:
some accepted perceived quality metric run over the above traces.
e.g., MOS, PESQ, NIQA, G.107, R-Factors, E-model?
}



}

\abbr{	
\subsection{Send-Side Prioritization}

\begin{figure}[tbp]
\centering
\includegraphics[width=0.49\textwidth]{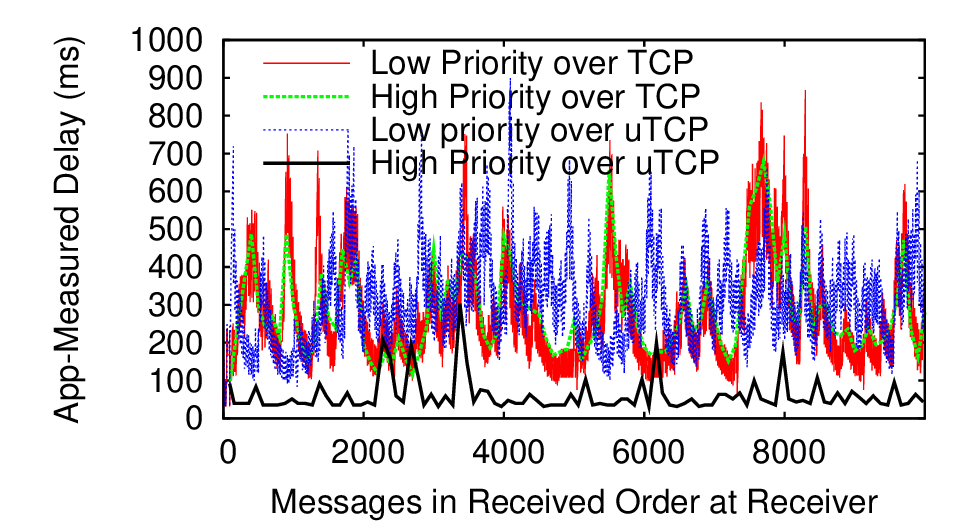}
\caption{Prioritized messages experience lower end-to-end delay with \utcp.}
\label{f:sendfunc}
\end{figure}

To test \utcp's sender-side prioritization,
we use a synthetic application that continuously sends
messages to the receiver at network-limited rate,
of which one in every 100 messages are considered ``high-priority.''
Figure~\ref{f:sendfunc}
plots application-observed messages delay over time,
for high- and low-priority messages,
atop TCP versus \utcp.
As expected,
high-priority messages consistently observe
much lower delays under \utcp
because they short-cut the TCP send queue.
The next section explores a more realistic application
for prioritization.

\xxx{	Compare send-buffer prioritization against
	send buffer auto-tuning mechanisms as a baseline.}

}{	
}

\subsection{VPN Tunneling}

Applications running atop TCP-based VPN tunnels
often encounter {\em TCP-in-TCP} effects~\cite{titz01why}.
The applications' tunneled TCP flows assume they are running
atop a best-effort, packet-switched network as usual,
but are in fact running atop a reliable, in-order TCP-based tunnel.
The TCP tunnel affects the tunneled flows' congestion control
by increasing observed latency and RTT variance,
and masks losses:
tunneled flows never see ``lost'' or ``reordered'' TCP segments,
only long-delayed ones.
While \utcp does not change TCP's reliability or congestion control,
it offers tunneled flows
lower delay and jitter,
and a more accurate view of packet losses.

To test Minion's impact on TCP-in-TCP effects,
we made two changes to OpenVPN 2.1.4~\cite{openvpn}.
First, 
we modified OpenVPN to use \ucobs instead of TCP,
enabling unordered delivery of tunneled IP packets.
\com{
OpenVPN's framing naturally works well with unordered delivery underneath,
and the delivering the tunneled IP datagrams out-of-order
only extends IP's unordered delivery properties through the tunnel.
}
Second,
to reduce delay variance of tunneled TCP flows further,
the modified OpenVPN gives tunneled TCP ACKs
a higher priority than other packets.

The experiment uses
a link with $3$Mbps download and $0.5$Mbps upload bandwidth,
consistent with the median speed of 
residential connections~\cite{internet-speed}. 

\begin{figure}[tbp]
\centering
\includegraphics[width=0.49\textwidth]{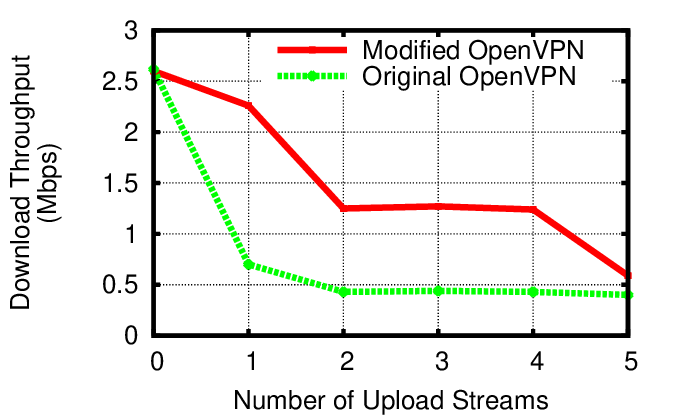}
\caption{Throughput obtained by a TCP flow through modified and unmodified OpenVPN.}
\label{f:ackpri}
\end{figure}

Figure~\ref{f:ackpri}
shows measured throughput of 
a single {\em download},
with original and modified OpenVPN tunnels,
for a varying number of competing {\em uploads}
within the same tunnel.
While using \utcp does not eliminate all TCP-in-TCP effects,
the reduced RTT and RTT-variance
noticeably improve performance.

\begin{figure}[tbp]
\centering
\includegraphics[width=0.49\textwidth]{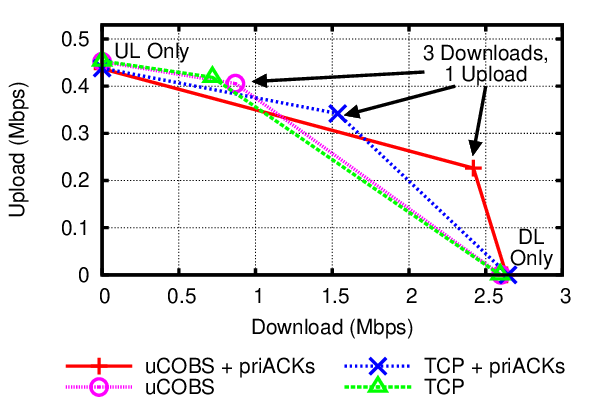}
\caption{Contribution of independent modifications to network utilization.}
\label{f:scatplot}
\end{figure}

\com{
we measure the performance of 
a TCP connection representing both download and upload
through an unencrypted VPN tunnel 
that also carries 
competing TCP connections in the opposite direction.
ACKs for the download stream thus 
compete for bandwidth with data from the competing upstream flows 
and vice versa. To individually study the effects of 
mentioned changes, the experiment is conducted on four different 
VPN configurations -- TCP Tunnel with and without ACK prioritization and
\ucobs Tunnel, with and without ACK prioritization. 
}		

To understand these performance improvements further,
we now measure total network utilization
achieved by independently adding the two modifications---%
unordered delivery at the receiving ends of the tunnel (labeled {\em uCOBS}),
and ACK prioritization at the sending ends (labeled {\em priACKs})---%
leading to four variants of OpenVPN.
Figure~\ref{f:scatplot}
shows total upload and download throughputs
obtained by the VPN tunnel
in three different contexts:
one upload (labeled {\em UL}) within the tunnel,
one upload competing with three downloads within the tunnel,
and
one download (labeled {\em DL}) in the tunnel.

With no competing flows,
labeled {\em UL Only} and {\em DL Only} 
in Figure~\ref{f:scatplot},
all four variants perform similarly.
With multiple competing downloads,
out-of-order delivery improves download performance by a small amount,
but ACK prioritization
greatly improves download performance.
Upload throughput suffers, 
however,
as ACK prioritization is added.
This throughput degradation is atrributable
to the poor interaction between 
the small \verb|write()|s of
ACK packets being sent through the tunnel
and Linux's \verb|skbuff|-based congestion control
described in Section~\ref{sec:bw-cpu}.
Despite this degradation to upload throughput,
the area under the curve---%
representing total network utilization---%
remains highest with the fully modified tunnel.
In a future version of \utcp
that fixes this Linux-specific issue,
we expect the upload throughput to remain high
even with ack-prioritization,
and network utilization to reflect 
\utcp's benefits more clearly.


\subsection{Multistreaming Web Transfers}

\begin{figure*}[t]
\centering
\includegraphics[width=0.75\textwidth]{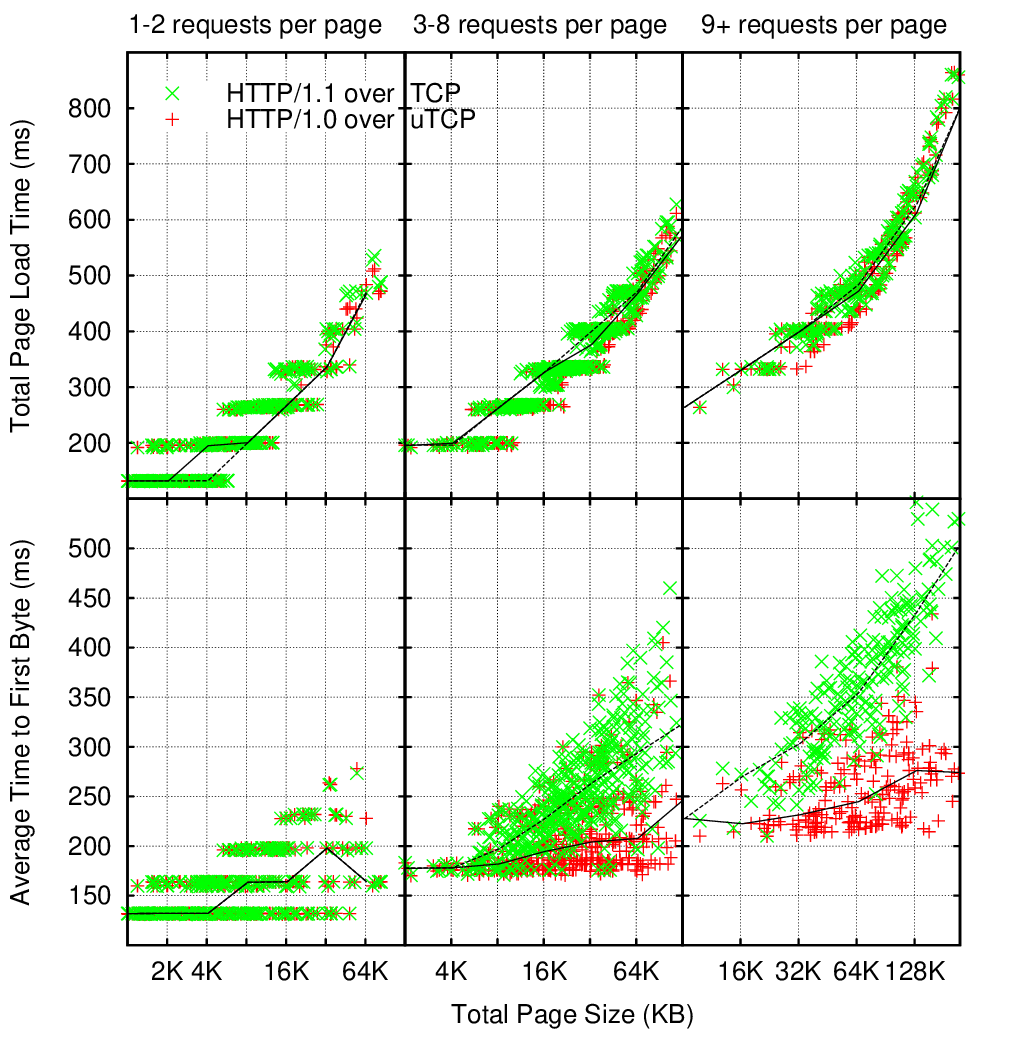}
\caption{Pipelined HTTP/1.1 over a persistent TCP connection, 
vs. Parallel HTTP/1.0 over \mstcp.}
\label{f:web}
\end{figure*}

\abbr{	
To explore \utcp's potential benefits for web browsing,
we built \mstcp,
a simple {\em multistreaming} protocol
providing multiple concurrent, ordered message streams 
atop a single \utcp connection.
While similar in purpose to SPDY~\cite{spdy},
to our knowledge \mstcp is the first TCP-based multistreaming protocol
that offers the unordered delivery benefits
of non-TCP-based multistreaming protocols
such as SCTP~\cite{rfc4960} and SST~\cite{ford07structured}.
We omit a detailed description of \mstcp for space reasons,
but its design follows standard techniques.

To evaluate \mstcp,
we compare the performance of
parallel HTTP/1.0-style object requests over \mstcp,
against pipelined HTTP/1.1 requests on a persistent TCP connection,
under a trace-driven web workload.
We use a fragment of the UC Berkeley Home IP web client traces
from the Internet Traffic Archive~\cite{ita},
using the trace to drive a series of web page downloads.
Each page consists of a ``primary'' request for the HTML, 
followed by ``secondary'' requests for 
embedded objects such as images. 
\com{
We sort the trace by client IP address so that 
each user's activities are contiguous, 
then use only the order and sizes of requests to drive the simulation, 
ignoring time stamps. 
Since the traces do not indicate which requests belong to one web page, 
the simulation approximates this information by 
classifying requests by extension into ``primary''
(e.g., '\verb|.html|' or no extension) 
and ``secondary'' 
(e.g., '\verb|gif|', '\verb|.jpg|', '\verb|.class|), 
and then associating each contiguous run of secondary requests 
with the immediately preceding primary request. 
}
The simulation pessimistically assumes that 
the browser cannot begin requesting secondary objects 
until it has downloaded the primary object completely, 
but at this point it can request 
all secondary objects in parallel. 
The experimental setup
uses a link with $1.5$Mbps bandwidth in each direction
and with a $60$ms RTT.

Figure~\ref{f:web}
shows 
a scatter-plot of total page load time in the top three graphs,
and in the bottom three graphs,
average time to load the first byte of an object in each page---%
the time at which the browser can potentially start rendering the object.
The dark curves show median times,
computed across buckets of web page sizes.
As the figure shows,
\mstcp does not affect total page load times noticeably.
\mstcp shows much lower delay
in {\em starting} to load many objects,
however,
since \mstcp interleaves different objects' chunks
within the persistent connection.

Figure~\ref{f:web}
shows the end-to-end impact on web browsing 
of \mstcp's application-level message chunking and multiplexing
in addition to the benefits of \utcp's out-of-order delivery.
These latency savings,
while not solely due to \utcp,
represent the potential savings when
web frameworks like SPDY~\cite{spdy} 
use \utcp,
and make HTTP/HTTPS more usable as a general purpose 
substrate for deploying 
latency-sensitive applications~\cite{popa10http}.
}{	

We now explore building concurrency
atop \utcp's unordered message service;
we build a {\em multistreaming} abstraction that
provides multiple independent and ordered message streams 
{\em within}
a single \utcp connection.
While both the need for and the benefits of 
concurrency at the transport layer have been
well known~\cite{smux, rfc4960, ford07structured, natarajan06sctp},
to our knowledge,
this is the first time
that true multistreaming has been built using TCP.

\com{
The multistreaming library
provides the abstraction of {\em streams}
Each application message is transmitted as part of a {\em stream}---%
there can be several such streams within 
a connection---%
such that data within a stream is delivered in-order,
and no order is maintained across different streams within a connection.
}

Our prototype implementation of 
a multistreaming library in user-space atop \utcp,
which we call {\em Multistreamed TCP} or \mstcp,
uses a $16$-bit stream identifier,
and a $16$-bit stream sequence number 
to order chunks within a stream.
An application specifies a stream identifier when using
\mstcp's {\tt ms\_sendmsg()} API;
\mstcp then breaks every application message down
into a series of {\em chunks}---%
the \mstcp-defined smallest multiplexable unit of data
within a connection.
Chunks are prepended with a $16$-byte chunk header,
which
includes the stream identifier for the message
and the stream sequence number,
and then finally transmitted using \ucobs.
At the receiver,
\mstcp's {\tt ms\_recvmsg()} 
receives unordered chunks from the underlying \ucobs,
parses the chunk header,
and adds each chunk to its stream's {\em stream-queue}---%
{\tt ms\_recvmsg()} maintains one queue for each stream identifier
that it encounters in the received chunks---%
the chunks are then delivered to the application in-order
within their respective streams.
Note that \mstcp's API can be trivially extended
to using \utls as well.

We use the Web as an
example application
to illustrate the behavior of \mstcp.
HTTP/1.1 addressed the inefficiency of short-lived TCP streams through
pipelining over 
persistent connections. 
Since \mstcp
attempts to offer the benefits of persistent streams 
with the simplicity of the one-transaction-per-stream model,
We now compare
the performance of
HTTP/1.0 with parallel object requests over \mstcp
against the behavior of
pipelined HTTP/1.1 over a persistent TCP connection, 
under a simulated web workload.

For this test 
we simulate a series of web page loads, 
each page consisting of a ``primary'' HTTP request for the HTML, 
followed by a batch of ``secondary'' requests for 
embedded objects such as images. 
As the simulation's workload
we use a fragment of the UC Berkeley Home IP web client traces 
available from the Internet Traffic Archive~\cite{ita}. 
We sort the trace by client IP address so that 
each user's activities are contiguous, 
then we use only the order and sizes of requests to drive the simulation, 
ignoring time stamps. 
Since the traces do not indicate which requests belong to one web page, 
the simulation approximates this information by 
classifying requests by extension into ``primary''
(e.g., '{\tt .html}' or no extension) 
and ``secondary'' 
(e.g., '{\tt gif}', '{\tt .jpg}', '{\tt .class}), 
and then associating each contiguous run of secondary requests 
with the immediately preceding primary request. 
The simulation pessimistically assumes that 
the browser cannot begin requesting secondary objects 
until it has downloaded the primary object completely, 
but at this point it can in theory request 
all of the secondary objects in parallel. 
The experimental setup
uses a link with $1.5$Mbps bandwidth in each direction
and with a $60$ms RTT,
typical of browsing over a residential connection.

Figure~\ref{f:web}
shows 
a scatter-plot of
total time to load the entire page
in the top three graphs,
and the 
bottom three graphs show
the average time to load the first byte of an object within each page---%
the expected time for an object to start being rendered at the browser.
The X-axis represents 
total webpage size.
The dark curves show median times, computed across webpages 
in log-sized buckets of webpage size---%
the solid curve shows the median for HTTP/1.1 over TCP
and the dashed curve for HTTP/1.0 over multistreaming.

The total time to transfer the pages
remains the same;
\mstcp's
bit-overheads do not affect
application-observed throughput
noticeably.
\mstcp however, 
shows much lower latency
in loading the first byte of objects,
since the objects' chunks
are interleaved within the persistent connection 
by \mstcp.

We note that the results
in Figure~\ref{f:web}
show the end-to-end impact on web browsing 
of \mssctp's application-level message chunking and multiplexing
in addition to the benefits of \utcp's out-of-order delivery.
These latency savings
thus represent the potential savings when
web frameworks like SPDY~\cite{spdy} 
use \utcp,
and make HTTP/HTTPS more usable as a general purpose 
substrate for deploying 
latency-sensitive applications~\cite{popa10http}.

}

\subsection{Implementation Complexity}

\begin{table}[tbp]
\begin{small}
\begin{center}
\begin{tabular}{l|cc|c|c|}
		& TCP	& \utcp			& DCCP	& SCTP	\\
\hline
Kernel Code	& 12,982 & {\bf 565} (4.6\%)	& 6,338	& 19,312 \\
\end{tabular}
\end{center}

\begin{center}
\begin{tabular}{l|c|ccc|}
		& \ucobs	& SSL/TLS	& \utls		& DTLS \\
\hline
User Code	& {\bf 732}	& 31,359	& {\bf 586} (1.9\%) & 4,734 \\
\end{tabular}
\end{center}
\end{small}

\caption{Code size of \utcp prototype
	as a delta to Linux's TCP stack,
	the \ucobs library,
	and \utls as a delta to {\tt libssl} from OpenSSL\@.
	Code sizes of ``native'' out-of-order transports
	are included for comparison.}
\label{t:codesize}
\end{table}

To evaluate the implementation complexity
of \utcp and the related application-level code,
Table~\ref{t:codesize}
summarizes the source code changes \utcp makes to Linux's TCP stack
in lines of code~\cite{cloc-153},
the size of the standalone \ucobs library,
and the changes \utls makes to OpenSSL's \verb|libssl| library.
The SSL/TLS total does not include
OpenSSL's \verb|libcrypt| library,
which \verb|libssl| requires but \utls does not modify.

With only a $600$-line change to the Linux kernel
and less than $1400$ lines of user-space support code,
\utcp provides a delivery service comparable to
Linux's $6,300$-line native DCCP stack,
while providing greater network compatibility.
In user space, \utls represents less than a $600$-line change
to the stream-oriented SSL/TLS protocol,
contrasting with OpenSSL's $4,700$-line implementation of DTLS,
which runs only atop out-of-order transports such as UDP or DCCP.

\section{Related Work}
\label{sec:related}


{\em New transports for latency-sensitive apps:}
Brosh et al.~\cite{brosh10delay} model TCP latency,
and identify the regions of operation for latency-sensitive apps with TCP\@.
While some of the considerations apply,
such as latency induced by TCP congestion control,
\utcp extends the working region for such apps by eliminating 
delays at the receiver.
\com{	out-of-place in related work section.
We will next be investigating elimination of sender-side delays
with \utcp,
to further extend the working region of the TCP protocol.
}

DCCP~\cite{kohler06dccp, rfc4340}
provides an unreliable, unordered datagram service
with negotiable congestion control.
SCTP~\cite{rfc4960}
provides unordered and partially-ordered delivery services
to the application.
Both DCCP and SCTP
face large deployment barriers on today's Internet, however,
and are thus not widely used.
\com{
As we show in Section~\ref{sec:services},
\utcp-based transports do not suffer from this drawback,
and can provide abridged versions of these transports,
with arguably the most important features preserved.
}

New transports such as SST~\cite{ford07structured}
and CUSP~\cite{terpstra10channel}
run atop UDP to increase deployability,
and UDP tunneling schemes have been proposed
for standardized Internet transports as well~\cite{
	tuexen10udp,phelan10dccp}.
Many Internet paths block UDP traffic as well, however,
as evidenced by the shift of popular VoIP applications
such as Skype~\cite{baset06analysis}
and VPNs such as DirectAccess~\cite{davies09directaccess}
toward tunneling atop TCP instead of UDP,
despite the performance disadvantages.

{\em Message Framing over TCP: }
Protocols such as 
HTTP~\cite{rfc2616}, SIP~\cite{rfc3261}, and iSCSI~\cite{rfc3720}, 
can all benefit from out-of-order delivery,
but use TCP for legacy and network compatibility reasons.
All use simple type-length-value (TLV) encodings,
which do not directly support out-of-order delivery even with \utcp,
because they offer no reliable way to distinguish a record header
from data in a TCP stream fragment.
While COBS~\cite{cheshire97consistent}
represents an attractive set of characteristics
for framing records to enable out-of-order delivery,
other encodings such as BABS~\cite{cardoso07bandwidth}
also represent viable alternatives.

\com{	I think these distract from the paper's may point,
	and they sound like claims that might invite knee-jerk reactions.
	- baf

{\em Congestion Sharing: }
TCP-Session~\cite{padmanabhan98addressing}
and Ensemble TCP~\cite{eggert00effects} 
seek to share congestion control state among multiple app streams.
While not a primary goal of our work,
\ucobs and \utls can offer some version of the same capability
as discussed in Section~\ref{sec:cm}.

MPTCP~\cite{ford10tcp},
proposed as an extension to TCP,
uses multiple TCP flows to achieve multipath.
One of MPTCP's explicit, primary goals
is to offer compatibility with middleboxes~\cites{ford11architectural}.
Network-compatible multipath can be achieved with 
\utcp-based transports, as we discuss in Section~\ref{sec:services},
and can be done with minimal changes to the kernel.
The cost however,
is that with PM-MPTCP, 
there are  multiple levels of reliability and buffering
that can be avoided under native MPTCP.

}

\xxx{ Send-buffer Auto-tuning:
	\cite{semke98automatic,goel02supporting}
}

\section{Conclusion}
\label{sec:concl}

For better or worse, TCP remains the most common substrate
for application-level protocols and frameworks,
many of which can benefit from unordered delivery.
Minion demonstrates that it is possible to obtain
unordered delivery from wire-compatible TCP and TLS streams
with surprisingly small changes to TCP stacks
and application-level code.
Without discounting the value of UDP and newer OS-level transports,
Minion offers a more conservative path toward
the performance benefits of unordered delivery,
which we expect to be useful to applications
that use TCP for a variety of pragmatic reasons.

\com{
All of the Internet transports designed since TCP,
despite their diverse characteristics,
embody a common recognition that many important applications
can benefit from out-of-order delivery.
None of these out-of-order transports except UDP, however,
has surmounted the high barriers to entry
that today's Internet effectively places on new protocols layered atop IP.
Even applications such as VoIP that traditionally run on UDP
are shifting to TCP tunneling for network compatibility reasons.
Instead of ignoring or fighting this trend,
we have demonstrated a small suite of protocols
that can offer applications out-of-order delivery
while maintaining strict wire-compatibility with TCP and even TLS.
These protocols offer latency-sensitive applications such as VoIP
performance benefits comparable to UDP or DCCP,
with the compatibility benefits of TCP and TLS.
As a bonus, the source code changes required
to common TCP and TLS implementations
to support this simple out-of-order delivery model
are an order of magnitude smaller than the implementations
of new transport protocols dedicated to offering the same functionality.
While we find these results promising,
further performance improvements are likely to be possible
by combining \utcp with other TCP stack improvements,
such as support for multistreaming,
negotiation of application-appropriate congestion control schemes,
or mechanisms to minimize delay introduced
by TCP's sender-side buffers.
}

\subsection*{Acknowledgments}

We wish to thank Jeff Wise, Dishant Ailawadi, Stuart Cheshire, Matt Mathis,
Kun Tan, and the anonymous reviewers for their valuable feedback
and contributions to early drafts.
This research was sponsored by the NSF
under grants CNS-0916413 and CNS-0916678.


\bibliography{os,net}

\begin{thebibliography}{10}

\bibitem{openarena}
{OpenArena} project.
\newblock \url{http://openarena.ws/}.

\bibitem{spdy}
{SPDY: An Experimental Protocol For a Faster Web}.
\newblock \url{http://www.chromium.org/spdy/spdy-whitepaper}.

\bibitem{zeromq}
{ZeroMQ}: The intelligent transport layer.
\newblock \url{http://www.zeromq.org}.

\bibitem{itu03transtime}
\abbr{ITU}{International Telecommunication Union}.
\newblock Recommendation {G.114}: One-way transmission time, May 2003.

\bibitem{rfc4787}
F.~{Audet, ed.} and C.~Jennings.
\newblock Network address translation {(NAT)} behavioral requirements for
  unicast {UDP}, Jan. 2007.
\newblock RFC 4787.

\bibitem{balakrishnan99integrated}
H.~Balakrishnan, H.~S. Rahul, and S.~Seshan.
\newblock An integrated congestion management architecture for {Internet}
  hosts.
\newblock In {\em \bibbrev{SIGCOMM}{ACM SIGCOMM}}, Sept. 1999.

\bibitem{baset06analysis}
S.~A. Baset and H.~Schulzrinne.
\newblock An analysis of the {Skype} peer-to-peer {Internet} telephony
  protocol.
\newblock In {\em \bibbrev{INFOCOM}{IEEE INFOCOM}}, Apr. 2006.

\bibitem{brosh10delay}
E.~Brosh, S.~A. Baset, V.~Misra, D.~Rubenstein, and H.~Schulzrinne.
\newblock The delay-friendliness of {TCP} for real-time traffic.
\newblock {\em IEEE Transactions on Networking}, 18(5):1478--1491, 2010.

\bibitem{carbone10dummynet}
M.~Carbone and L.~Rizzo.
\newblock {Dummynet Revisited}.
\newblock {\em {ACM CCR}}, 40(2), Apr. 2010.

\bibitem{cardoso07bandwidth}
J.~S. Cardoso.
\newblock Bandwidth-efficient byte stuffing.
\newblock In {\em IEEE ICC 2007}, 2007.

\bibitem{rfc3234}
B.~Carpenter and S.~Brim.
\newblock {Middleboxes: Taxonomy and Issues}, Feb. 2002.
\newblock RFC 3234.

\bibitem{cheshire97consistent}
S.~Cheshire and M.~Baker.
\newblock {Consistent Overhead Byte Stuffing}.
\newblock In {\em ACM SIGCOMM}, Sept. 1997.

\bibitem{cisco-rbscp}
Cisco.
\newblock {Rate-Based Satellite Control Protocol}, 2006.

\bibitem{clark90architectural}
D.~D. Clark and D.~L. Tennenhouse.
\newblock Architectural considerations for a new generation of protocols.
\newblock In {\em \bibbrev{SIGCOMM}{ACM SIGCOMM}}, pages 200--208, 1990.

\bibitem{cloc-153}
A.~Danial.
\newblock {Counting Lines of Code, ver. 1.53}.
\newblock \url{http://cloc.sourceforge.net/}.

\bibitem{davies09directaccess}
J.~Davies.
\newblock {DirectAccess} and the thin edge network.
\newblock {\em Microsoft TechNet Magazine}, May 2009.

\bibitem{rfc5246}
T.~Dierks and E.~Rescorla.
\newblock The transport layer security {(TLS)} protocol version 1.2, Aug. 2008.
\newblock RFC 5246.

\bibitem{rfc2616}
R.~Fielding et~al.
\newblock Hypertext transfer protocol --- {HTTP/1.1}, June 1999.
\newblock RFC 2616.

\bibitem{ford07structured}
B.~Ford.
\newblock Structured streams: a new transport abstraction.
\newblock In {\em \bibbrev{SIGCOMM}{ACM SIGCOMM}}, Aug. 2007.

\bibitem{ford08breaking}
B.~Ford and J.~Iyengar.
\newblock Breaking up the transport logjam.
\newblock In {\em \bibbrev{HotNets-VII}{7th Workshop on Hot Topics in Networks
  (HotNets-VII)}}, Oct. 2008.

\bibitem{ford09efficient}
B.~Ford and J.~Iyengar.
\newblock Efficient cross-layer negotiation.
\newblock In {\em \bibbrev{HotNets-VIII}{8th Workshop on Hot Topics in Networks
  (HotNets-VIII)}}, Oct. 2009.

\bibitem{rfc5382}
S.~{Guha, Ed.}, K.~Biswas, B.~Ford, S.~Sivakumar, and P.~Srisuresh.
\newblock {NAT} behavioral requirements for {TCP}, Oct. 2008.
\newblock RFC 5382.

\bibitem{guo06delving}
L.~Guo, E.~Tan, S.~Chen, Z.~Xiao, O.~Spatscheck, and X.~Zhang.
\newblock {Delving into Internet Streaming Media Delivery: a Quality and
  Resource Utilization Perspective}.
\newblock In {\em {IMC}}, Oct. 2006.

\bibitem{honda11possible}
M.~Honda, Y.~Nishida, C.~Raiciu, A.~Greenhalgh, M.~Handley, and H.~Tokuda.
\newblock Is it still possible to extend {TCP}?
\newblock In {\em \bibbrev{IMC}{Internet Measurement Conference}}, Nov. 2011.

\bibitem{ita}
The {Internet} traffic archive.
\newblock \url{http://ita.ee.lbl.gov/}.

\bibitem{iyengar12minion-full}
J.~Iyengar, S.~O. Amin, M.~F. Nowlan, N.~Tiwari, and B.~Ford.
\newblock Minion: Unordered delivery wire-compatible with {TCP} and {TLS} (full
  version), Apr. 2012.
\newblock Technical Report. \texttt{arXiv:1103.0463}.

\bibitem{rfc4301}
S.~Kent and K.~Seo.
\newblock Security architecture for the {Internet} protocol, Dec. 2005.
\newblock RFC 4301.

\bibitem{rfc4340}
E.~Kohler, M.~Handley, and S.~Floyd.
\newblock Datagram congestion control protocol {(DCCP)}, Mar. 2006.
\newblock RFC 4340.

\bibitem{kohler06dccp}
E.~Kohler, M.~Handley, and S.~Floyd.
\newblock Designing {DCCP}: Congestion control without reliability.
\newblock In {\em \bibbrev{SIGCOMM}{ACM SIGCOMM}}, 2006.

\bibitem{mogul03tcp}
J.~Mogul.
\newblock {TCP} offload is a dumb idea whose time has come.
\newblock In {\em HotOS IX}, May 2003.

\bibitem{natarajan06sctp}
P.~Natarajan et~al.
\newblock {SCTP}: An innovative transport layer protocol for the {Web}.
\newblock In {\em \bibconf[15th]{WWW}{World Wide Web Conference}}, May 2006.

\bibitem{itu07wideband}
I.-T. T. S.~S. of~ITU.
\newblock Wideband extension to recommendation p.862 for the assessment of
  wideband telephone networks and speech codecs, Nov. 2007.

\bibitem{openssl}
{OpenSSL} project.
\newblock \url{http://www.openssl.org/}.

\bibitem{openvpn}
{OpenVPN} project.
\newblock \url{http://openvpn.net/}.

\bibitem{phelan10dccp}
T.~Phelan.
\newblock {DCCP Encapsulation in UDP for NAT Traversal (DCCP-UDP)}, Aug. 2010.
\newblock Internet-Draft draft-ietf-dccp-udpencap-02 (Work in Progress).

\bibitem{popa10http}
L.~Popa, A.~Ghodsi, and I.~Stoica.
\newblock {HTTP} as the narrow waist of the future {Internet}.
\newblock In {\em \bibbrev{HotNets-IX}{9th ACM Workshop on Hot Topics in
  Networks (HotNets-IX)}}, Oct. 2010.

\bibitem{rfc768}
J.~Postel.
\newblock User datagram protocol, Aug. 1980.
\newblock RFC 768.

\bibitem{reis08detecting}
C.~Reis et~al.
\newblock Detecting in-flight page changes with web tripwires.
\newblock In {\em \bibconf[5th]{NSDI}{Symposium on Networked System Design and
  Implementation}}, Apr. 2008.

\bibitem{rfc2818}
E.~Rescorla.
\newblock {HTTP} over {TLS}, May 2000.
\newblock RFC 2818.

\bibitem{rfc4347}
E.~Rescorla and N.~Modadugu.
\newblock Datagram transport layer security, Apr. 2006.
\newblock RFC 4347.

\bibitem{rosenberg08udp}
J.~Rosenberg.
\newblock {UDP} and {TCP} as the new waist of the {Internet} hourglass, Feb.
  2008.
\newblock Internet-Draft (Work in Progress).

\bibitem{rfc3261}
J.~Rosenberg et~al.
\newblock {SIP:} session initiation protocol, June 2002.
\newblock RFC 3261.

\bibitem{rfc3720}
J.~Satran, K.~Meth, C.~Sapuntzakis, M.~Chadalapaka, and E.~Zeidner.
\newblock Internet small computer systems interface {(iSCSI)}, Apr. 2004.
\newblock RFC 3720.

\bibitem{rfc3550}
H.~Schulzrinne, S.~Casner, R.~Frederick, and V.~Jacobson.
\newblock {RTP}: A transport protocol for real-time applications, July 2003.
\newblock RFC 3550.

\bibitem{rfc4960}
R.~{Stewart, ed.}
\newblock Stream control transmission protocol, Sept. 2007.
\newblock RFC 4960.

\bibitem{rfc793}
Transmission control protocol, Sept. 1981.
\newblock RFC 793.

\bibitem{terpstra10channel}
W.~W. Terpstra, C.~Leng, M.~Lehn, and A.~P. Buchmann.
\newblock Channel-based unidirectional stream protocol {(CUSP)}.
\newblock In {\em \bibbrev{INFOCOM}{IEEE INFOCOM} Mini Conference}, Mar. 2010.

\bibitem{titz01why}
O.~Titz.
\newblock Why {TCP} over {TCP} is a bad idea, Apr. 2001.
\newblock \url{http://sites.inka.de/bigred/devel/tcp-tcp.html}.

\bibitem{tuexen10udp}
M.~Tuexen and R.~Stewart.
\newblock {UDP Encapsulation of SCTP Packets}, Jan. 2010.
\newblock Internet-Draft draft-tuexen-sctp-udp-encaps-05 (Work in Progress).

\bibitem{speex}
J.-M. Valin.
\newblock The speex codec manual version 1.2 beta 3, Dec. 2007.
\newblock \url{http://www.speex.org/}.

\bibitem{rfc908}
D.~Velten, R.~Hinden, and J.~Sax.
\newblock Reliable data protocol, July 1984.
\newblock RFC 908.

\bibitem{websocket}
W3C.
\newblock The websocket api (draft), 2011.
\newblock \url{http://dev.w3.org/html5/websockets/}.

\bibitem{internet-speed}
www.speedmatters.org.
\newblock 2010 report on internet speeds in all 50 states, Nov. 2010.
\newblock \url{http://www.speedmatters.org/content/internet-speed-report}.

\bibitem{zec02estimating}
M.~Zec, M.~Mikuc, and M.~Zagar.
\newblock Estimating the impact of interrupt coalescing delays on steady state
  {TCP} throughput.
\newblock In {\em SoftCOM}, 2002.

\end{thebibliography}
\bibliographystyle{abbr}

\end{document}